\documentclass{elsart}

\usepackage{amsmath, amscd, amssymb}
\usepackage{mathrsfs}

\journal{Physica A}

\newcommand{\F}{{\Phi}}

\newcommand{\HH}{{\mathscr{H}}}

\newcommand{\rhs}{\F\subset\HH\subset\F^\times}

\newcommand{\BH}{\mathbb H}

\newcommand{\R}{{\mathbb R}}

\newcommand{\II}{{\mathbb I}}
\newcommand{\C}{{\mathbb C}}

\newcommand{\CS}{{\mathscr{S}}}

\newcommand{\g}{{\mathfrak{g}}}
\newcommand{\hh}{{\mathfrak{h}}}

\newcommand{\bee}{\begin{enumerate}}
\newcommand{\ene}{\end{enumerate}}
\newcommand{\een}{\end{enumerate}}
\newcommand{\bea}{\begin{eqnarray}}
\newcommand{\eea}{\end{eqnarray}}
\newcommand{\beq}{\protect{\begin{equation}}}
\newcommand{\eeq}{\protect{\end{equation}}}

\begin{document}

\begin{frontmatter}

\title{{Generalized Wavefunctions for \\
Correlated Quantum Oscillators IV: \\
Bosonic and Fermionic Gauge Fields.}}

\author{S. Maxson}
\address{Department of Physics\\University of Colorado
Denver\\Denver, Colorado 80217\\ }
\ead{steven.maxson@ucdenver.edu}

\date{\today}

\begin{abstract}

The hamiltonian quantum dynamical structures in rigged Hilbert 
spaces used in the preceding three installments to represent correlated hamiltonian 
dynamics on phase space are shown to possess a well defined 
covering structure, which is demonstrated explicitly in terms of Clifford
algebras.  The unitary Clifford algebras are described here for the first time, 
and arise from the intersection of the orthogonal and common symplectic 
(Weyl) Clifford algebras of the complexification of the canonical phase 
space. The convergence of the exponential map is possible in available 
topologies in our constructions, but it does not converge without additional 
assumptions in general.   Continuous dynamics exists only in semigroups. 
A well defined spin geometry exists for the 
unitary Clifford algebras in the appropriate Witt basis, which also affords us
both bosonic and fermionic representations through alternative
topological completions of the same structure, and physically
represent the stable states of the system.  Unitary
Clifford algebras can be used to define dynamical gauge
bundles for arbitrary numbers of correlated (unified)
fields.  The generic dynamical gauge group for four pairs of
canonical variables (four fields) is shown to be isomorphic to $U(4)\times U(4)$,
with the spectrum effectively determined by $S[U(4)\times U(3) \times U(1)]$ 
due to the constraint of
geodesic transport of the generators of the dynamical group.  It is conjectured
this is an unified version of $U(4)$ gauge gravity wherein particles
correspond to islands of dynamical stability. An
isomorphism is shown explicitly demonstrating the ability to associate these
structures over four pairs of canonical variables with covariant structures in a
non-trivial spacetime with $(+,-,-,-)$ local signature.  The
covariance of the identity of elementary particles follows, and is of
dynamical origin, and by inference PCT is of dynamical origin also.  
An area of fundamental conflict is demonstrated between
notions of noncompact hamiltonian dynamics and general covariance, and
a resolution is proposed for the present constructions. This includes prediction
of the existence of chimeric bosons, whose quantum numbers are
not covariant so they may appear to have a different identity to
different observers.
\end{abstract}

\begin{keyword}{unitary Clifford algebra  \sep dynamical gauge fields \sep non-covariant bosons}

\PACS   11.15.-q \sep 11.15.Tk \sep 03.65.Fd \sep 02.10.De

\end{keyword}

\end{frontmatter}

\section{Introduction}\label{sec:intro}

In this fourth and concluding installment concerning
correlations of quantum oscillators, we relate what has gone in the
preceding installments to an overall mathematical structure which is
(reasonably) well defined mathematically.  Additionally, we will show
ways in which a correlated, or
unified, hamiltonian quantum field theory over canonical variables can be
associated to extended objects in a curved relativistic spacetime.
When a field theoretic interpretation is adopted for our
constructions, there are some
interesting generic implications for quantum field theory and
particle physics, and we will demonstrate a dynamical fiber bundle
structure of physical interest existing within our hamiltonian dynamics 
formalism.  We will use
this as the basis for a gauge field theory, but
will not, however, make any effort in this first description
to present anything like a mature theory of particle physics, and
will argue extensively from analogy to the Standard Model.  Our methods
are largely generic to any probabilistic representation of correlated
hamiltonian dynamics, and it is interesting to see confirmation of the
gauge group of the Standard Model in the generic gauge structure for three
fields.  Our generic methods predict the spectrum of the Standard Model exactly,
although there are differences in interpretive detail (which are mathematically
driven in the present case).

We likeswise make generic predictions of the gauge
group for four fields or any other number of fields, and, for four
fields a fairly straightforward generalization of the Standard Model
is indicated.  In addition to the gauge related particle spectrum,
from the mere existence of hyperbolic dynamics in
nature, we  make the prediction of additional chimeric
bosons which are ``fundamental'' but not ``elementary'', and these
may possibly offer an explanation for the mysteries of 
Dark Matter and Dark Energy.  These
are  characterized as chimeric since the quantum numbers which
characterize  them lack covariant associations, unlike the quantum
numbers  associated with the gauge group we deduce for four fields.
They may appear as different particles to different observers in
consequence, and one mechanism for this is shown.

This is a general probabilistic correlated dynamics, and we could, for instance, 
use this same general formalism to describe the dynamics of a fluid
with a given number of pairs of canonical variables, either as an
ensemble or as a field, which incorporates the presence of internal
correlation such as might result in the case of long range forces with
self-consistent interactions.  Whatever physical problem is being
represented in our formalism, the dynamical fiber bundle structure
gives us a representation of the islands of stability within a richer
and fuller description of dynamics which includes resonances, and
which is also significantly more complex mathematically.  We further
will demonstrate the mathematical and physical importance of
understanding that our constructions are in fact spin
constructions, and show that it is
extremely useful to regard our constructions as associated with
representation of new types of Clifford algebra, the unitary
Clifford algebras, existing only for spaces possessing both
symplectic and orthogonal structure, such as phase space.  A
Clifford algebra
provides the covering structure to make this work well behaved
mathematically, and our generalized wave functions are a part of a
representation of this Clifford algebra of phase space and its
complexification.

Spinors are associated with the Clifford algebras and their
representations, and there are many subtleties we shall gloss and take an
optimistic view of in the present forum.  Spinor structures and spin
geometry can be problematic, but we have
grounds for feeling secure with respect to the key elements (the
unitary Clifford algebras) we depend on.  We would also suggest that
the unitary Clifford algebras provide a well defined nucleus which
may be extended to a full symplectic Clifford algebraic structures
without arbitrary conditions imposed to insure convergence of the
exponential map (as is the case for the symplectic formal series type
of Clifford algebra~\cite{crumeyrolle}).  There appear to be
symplectic Clifford algebras containing
well defined semigroups of symplectic (=dynamical) transformations,
even though full groups do not seem available without additional,
probably arbitrary, assumptions.  This clearly implies, as the physical
interpretation of the mathematics we were compelled to adopt,
that irreversibility is intrinsic to dynamics, a long held contention of the
late Professor Prigogine: it appears that, from the structure of phase space
itself, one can deduce continuous dynamics only for the case when 
correlation exists within the dynamical system, and that this continuous dynamics is
irreversible, in general being associated with semigroups only, unless one adds
essentially arbitrary mathematical assumptions in order to obtain a group
structure and invertibility.  (The semigroups are obtained by topological 
completion of polynomials into the exponential map using the only topology
which is naturally present in our complexification of phase space itself, 
and this is a complex
hyperbolic topology, the hyperbolic Kobayashi semidistance topology.  I know of no
inequivalent topology in which convergence of continuous dynamical
transformations is well defined without arbitrary assumptions--but I may be
insufficiently clever to have found such a topology, so will
assert only an implication and not a well proven result.)

As to this nucleus, we have a sufficiently well defined spin structure
to possess a well defined Yang-Mills (principal fiber or
gauge bundle) structure associated to it.  We
will restrict ourselves to those dynamical aspects of the spin geometry
issues of concern to field theorists, and leave broader dynamical
concerns for another day.  It is noteworthy, however, that formulating
our theory in a form capable of reflecting a dynamical arrow of time
(the semigroups of symplectic transformations provide a vehicle for
expressing the boundary and
initial conditions of an irreversible dynamical process) is
mathematically sufficient to make our hamiltonian quantum dynamical
theory well defined. Whether such a semigroup formulation is also
necessary seems to touch on areas of great subtlety and complexity,
and will addressed in some detail, but not entirely resolved.

As indicated in installment one~\cite{I}, our approach admits a
field theoretic interpretation.  If
``dynamics is the geometry of behavior'', we would add that the most
suitable description of geometry seems to be in terms of geometric
(Clifford) algebra.  The present paper emphasizes Clifford algebra
issues and field theoretic interpretation of Clifford algebra
representations for various Clifford algebras associated with phase
spaces. After dealing with some left over matters from the preceding
installments, we will establish the Lie algebra valued connection
constructively, demonstrate the existence of generalized Yang-Mills
gauge structures, and show how the basic gauge group of the
Electroweak Theory and Standard
Model, and also the canonical gauge structure for four fields, emerge
in a breathtaking and natural way, merely by looking at the canonical
transformations of appropriate numbers of oscillators (identified in
the usual manner with fields).  The elementary particles are the
islands of stability in this dynamical structure of the potentials of
the gauge fields.  The gauge groups are exact, however,
and there is no ``spontaneous symmetry breaking'' associated with their
definition~\cite{note0x5}.  The unitary
gauge group emerging from this unitary Clifford algebra approach
possesses both fermionic (even dimensional) and bosonic (odd
dimensional) representations, corresponding to whether we view $U(N)$
as a subgroup of an orthogonal group, lying in an orthogonal Clifford
algebra (fermionic) or as part of a symplectic (semi-)group, lying in
a symplectic (bosonic) Clifford algebra of some sort.  It is therefore
a mathematical error to mix bosons and fermions in our
hamiltonian formalism.  We suggest that there is some
correspondence between at least some parts of the even dimensional (fermionic) and odd
dimensional (bosonic) representations, since they do represent the same group.
This identification is probably related to
topological notions, but we will not attempt any detailed justification at
present.  We do wish, however, to clearly indicate that at the present
such terms as ``quark--gluon plasma'' have a very muddy meaning
mathematically. We will ignore all but the most superficial
representation issues.

As to the unitary Clifford algebras, the choice between
bosons and fermions is between alternative topological completions, and, seeing
that we can never conduct a Cauchy sequence of measurements, the two
alternative topological completions should be physically
indistinguishable--matrix elements should not change.  Bosons are
products of the symplectic geometry of phase space, while fermions
arise when you thinks of phase space from the perspective of
orthogonal geometry.  Seemingly, this is where inescapable
mathematical necessity
has led us in our pursuit of a well defined unified (correlated) hamiltonian
quantum field theory, and is either physically relevant or it isn't.  In any
event, the hamiltonian and Lagrangian approaches to field theory may lead to
significantly different end points~\cite{note0x7}.

There are a couple of key ingredients in our geometric structure
that play essential roles in our construction.  Spinors are usually
frame dependent (this is the reason there is no spinor calculus
analogous to the tensor calculus), and the unitary transformations will
leave our real Witt frames invariant (much like the orthogonal
transformations leave conventional frames invariant).  The real Witt
bases enable us to obtain a well defined (but frame dependent)
differential geometry for our  symplectic spinors from the well known
spin geometry of orthogonal
spinors (a special topic in Riemannian geometry~\cite{lawson}.)  There
is a simple mathematical trick using standard theorems of topology for
linear spaces which we invoke to obtain this result.  Secondly, we
have a weak symplectic form (see~\cite{II,III}), meaning that Darboux's
theorem does not apply, and our geometry can be other than locally
Euclidean, permitting the existence of non-trivial local invariants
such as curvature, which are prerequisite for a non-trivial gauge
theory.

A brief exercise will demonstrate that analytic continuation of the
traditional Hilbert space does not result in vectors possessing
Bose-Fermi symmetries which
are well defined.  This follows because the energy spectrum for
vectors belonging to that analytically continued
space is not necessarily bounded from below.  Let us
consider energy eigenvectors $\vert a\rangle$ and $\vert
b\rangle$ belonging to some space for which there is a well defined
``vacuum'' or minimum energy eigenvector, $\vert 0\rangle$.  Then
there is some transformation $A$ such that $\vert a\rangle = A \vert
0\rangle$ and some transformation $B$ such that $\vert b\rangle =B
\vert 0 \rangle$.  Without loss of generality we may regard $A$
and $B$ as esa, and it follows that
\begin{equation}
\langle a \vert b\rangle = \langle 0\vert \;\; \frac{ AB+BA}{2} \; +\;
\frac {AB-BA}{2} \;\; \vert 0\rangle
\label{eq:decomp}
\end{equation}
The uniqueness of this decompositon into symmetric and antisymmetric
parts depends on the existence of a unique fiducial vector, such as
$\vert 0\rangle$.  When the energy spectrum is unbounded below, there
is no such fiducial vector, and many similar decompositions can exist,
with nothing to distinguish any particular one.
This brief demonstration illustrates that the traditional form of the
boson-fermion superselection
rule does not apply to analytically continued systems, in which
the energy spectrum is not bounded from below.  However, for our
multicomponent state vectors there is a somewhat more complicated
situation than this naive calculation is relevant to, which we
elaborate in detail.

For the
multicomponent spinor formulation developed in preceding installments,
especially installment two~\cite{II}, and further specified below,
any bilinear form on
phase space must be either strictly symmetric or strictly antisymmetric.
(This is a characteristic of Clifford algebras in general.)
This compels us to choose one or the other bilinear form for the
construction of our Clifford algebra, although there is a special
basis for phase space compatible with both the ordinary orthogonal
(symmetric) form and the (antisymmetric) symplectic form.  In this special
basis, the real Witt basis, we can simultaneously generate
representations of either, enabling us to form the
non-trivial intersection of the
orthogonal and symplectic Clifford algebras of phase space.  The
unitary Clifford algebra which results thus has a canonical basis in
which one may alternatively consider physical aspects associated with the
orthogonal perspective, such as fermionic representations of bulk
matter by Dirac spinors in a spacetime with local signature
$(+,-,-,-,)$, or those aspects associated with the symplectic perspective,
such as dynamics, forces, interactions, etc., associated with bosons,
represented by symmetric spinors.  We will refer to such choices of
representation as a choice of perspective for our state vectors, and is
in some sense dependent on the choice of dimension for the representation:
even dimensional representations are fermionic and odd dimensional
representations are bosonic.  We can thus think of a system as a bunch of
fermions (particles) or as a bunch of bosons( intermediaries of the forces--the
dynamical entities), but must consider a particle as either fermionic ``lumps of
geometry'' or as bosonic ``lumps of dynamical fields''.  These perspectives are
alternative ways  of looking at one physical structure in alternative representations,
according to our constructions of those representations of physical
structures using a unitary Clifford algebra.  In this view,
the boson-fermion dichotomy is an artifact of the representation
chosen for the stable structures of the theory (such as particles or
other stable structures, as in, e.g., stable circulation of fluids),
and need not be an intrinsic property of the underlying structure 
being represented.  See also Section~\ref{sec:tentative} below, but note
the discussion at the end of this section.
The representation chosen will be associated, in turn, with a choice
of topological completion of an algebraic set, and topology is not an
experimental observable (you can never conduct a Cauchy sequence of
measurements). {\em The alternative perspectives, once adopted, may have
different observables in the physical context consistent with the underlying
principles of the perspective which has been adopted.}  Our  use of
the terms boson and fermion may not, in consequence,
exactly correspond to the usual conventions of quantum theory, but
they have mathematical precision.

The decomposition of equation (\ref{eq:decomp}) can be said to be
unique in a unitary Clifford algebra such as we construct in
Section~\ref{sec:spinors} in the sense that each of the two terms is
non-trivial in one perspective only, each perspective being associated
with cofactors over ideals based on one or the other of the
alternative bilinear
forms which exist separately on the space.  The two terms cannot mix to
define a mixed bilinear form on our spin-vectors, which form
a representation of phase space (as part of the representation of the
Clifford algebra of phase space).  In our constructions, bosons and
fermions are associated with separate and distinct,
unique bilinear forms, and each
bilinear form defines a perspective, but the perspectives (e.g.,
representations) may not
mix.  Thus, you may speak of the fermionic properties of bulk matter
or you may
speak of bosonic forces and dynamical evolution, but you must change
perspective between these two alternatives, and really cannot properly
talk of both {\em simultaneously} without exceeding the bounds of
mathematical propriety.  To consider the electrodynamic
interaction of two electrons, for instance, one must consider each
electron as a ``conglomeration of dynamical field stuff'' in order to speak of the
exchange of photons (other ``conglomerations of dynamical field stuff'')
between them.  {\em Dynamics} is the exclusive jurisdiction of the
perspective associated with the symplectic form and factorization of
the tensor algebra of phase space over that bilinear
form yields a Clifford algebra
suitable only for the representation of bosons (and that Clifford
algebra is
properly represented exclusively by odd dimensional symmetric
spinors).  There
is no superselection rule in our RHS spin formulation in the same
sense as such
a rule is applied to the conventional Hilbert space quantum theory.
Rather,
there is a selection between perspectives (Clifford algebras and
their representations).

All of the stable or quasi-stable states arising through symplectic
transforms in our hamiltonian dynamical formalism are extremely
closely related formally
to the coherent and squeezed states of the electromagnetic field so
well known to quantum optics.  Using photons and Foch space in the
usual formalism of quantum optics, for instance, there is limited
localizability--localizability is limited by the position-momentum
uncertainty principle.  The notion of position-momentum minimum
uncertainty comes from considering the area of an ellipse in phase
space using familiar notions of Euclidean geometry, i.e., stems from
an orthogonal (symmetric) metric (bilinear form), and not from the
symplectic (antisymmetric) bilinear form.  Our stable and quasi-stable
states are nothing more than squeezed (minimum uncertainty) states of
numerous fields, just as photons may represent squeezed states of
the single electromagnetic field. The transition in one's thinking
from the notion of bosonic squeezed state to notion of
fermionic minimum uncertainty state illustrates the subtlety
of the transition between perspectives in our constructions.  The
creation--annihilation operator formalism suggests that we are
operating on the ``particle'' side of wave-particle duality, although in
Section~\ref{sec:tentative} we shall indicate that such matters as
energy scale
really govern, just as they do in quantum optics and atom optics
(e.g., Compton vs. de Broglie wavelength, and so on).

In terms of
equation (\ref{eq:decomp}), we would say that there are fermionic and
bosonic representations of the operators $A$ and $B$, appropriate to
the two alternative perspectives, and one or the other of the two
mixed operator terms will vanish in a given perspective.  
All is predicated on our choice
of the Witt basis, yet another instance of the basis dependence of
spinors.  The scalar product in
equation (\ref{eq:decomp}) will survive as
either symmetric or skew depending on whether the orthogonal
(symmetric) or symplectic (skew) form is chosen for factorization of
our tensor algebra.  The strictly alternative representations are also
distinguishable from the thing being represented, since they are
isomorphic to alternative topological completions of a set,
and, in any event, the representation isomorphisms are
not natural isomorphisms so there is some
inequivalence aside from any topological issues.  (See, e.g.,
Section~\ref{sec:fermions} and reference~\cite{hanany} for examples
of inequivelance.)

\section{Necessity of Spinor Structures}\label{sec:necessity}

In the following two subsections, we pursue the reasons for use of
spinors in our representation of the correlated combinations of
oscillators problem.  The puzzling structure motivating this is the
conjugacy of the (complex) symplectic transformations between $iY$ and
$Z$ seen in installment two~\cite{II}.  Of course, spinors figure in group
representations, providing the ``fundamental representations'', and
there are some technical mathematical reasons that make them
appealing (even mandatory), but
there are strong physical reasons as well.  They make
our hamiltonian dynamical structure well defined.

It is  natural to avail ourselves of the spinoplectic
covering structure or perhaps
even to use the representation of the full Clifford algebra
itself.  This has the further virtue of making our representation
structure into Clifford modules, which are well known and well
studied~\cite{note7}.

Although we may have uncertainties about the best way to interpret
stability implications of the conjugacy of $iY$ and $Z$ seen in
installment two~\cite{II}, at least there is a covering structure in
which the conjugacy is well defined, without regard to the appearance
of an apparently undefined inverse semigroup transformation in it.
This is because we are working with spinors: our spaces of states
possess an orientation by virtue of their symplectic structure (as do
all spaces representing quantized systems according to accepted notions
of geometric quantization) and
$exp$ is holomorphic for us, so the first two Stieffel-Whitney classes
vanish, making our generalized spaces of quantum states spin spaces by
construction.  To insist that our group representations be UIR's
would be a grievous mathematical error.
Our representation spaces $\F_{\mathfrak{sp}(4,\R )^\C
  \pm}$ (and their function space realizations) possess a complex
symplectic structure (since their automorphism group is $Sp(4,\R
)^\C$.)  Spinors are ideals of Clifford algebras, so we conclude that
our representation space(s) is part of the representation of some sort of a
symplectic Clifford algebra~\cite{crumeyrolle}.

The spinors physicists are most familiar with arise in the
representation of orthogonal Clifford algebras, such as the Dirac
spinors.  Typical spinors of physics are associated with a Clifford
algebra for some space $V$ which possesses an orthogonal structure
(symmetric bilinear form, elliptic scalar product, etc.), denoted
$Cl_O(V)$.  This universal Clifford algebra for $V$ contains a group,
the (orthogonal) Clifford group $\mathscr{G}_O (V)$.  The (orthogonal)
Clifford group contains a spin group, $Spin (V)$, which in turn
provides a double cover for the group of orthogonal transformations on
$V$, $O(V)$.  If we represent an orthogonal trnaformation on $V$
belonging to $Cl_O(V)$ by an exponential, $( e^{iX\theta /2})_{O(V)} \in
O(V)$, then the orthogonal rotation of a vector $A \in V \subset
Cl_O(V)$ about the direction given by the vector $X$ is representated
as the conjugation
\begin{equation}
\left( e^{iX\theta /2} \right)_{O(V)} \, A \, \left( e^{-iX\theta /2}
  \right)_{O(V)} \quad .
\label{eq:orthrot}
\end{equation}
This exact same rotation is represented in the $Spin(V) \subset CL_O
(V)$ covering structure as
\begin{equation}
\left( e^{iX\theta} \right)_{Spin(V)} \, A \quad ,
\label{eq:spinrot}
\end{equation}
i.e., as an operation from the left to right, without conjugation.  In
other words, a semigroup orthogonal rotation which is performed by
conjugation (and therefore has only a conditional local meaning, at
best) determines a well defined spin transformation which acts from
the left only, therefore defining a unique geometric structure having
a global meaning on our space of (generalized) states.  We will show
{\em infra} that the unitary transformations form what is effectively
a group substructure within both the orthogonal and symplectic semigroups
of transformations.  The group structure of these unitary transformations
are shown below in many places and in many ways to be the mathematical
key to the well defined mathematical structures in our constructions.

There is a similar hierarchy of groups in the case of the symplectic
Clifford algebras~\cite{crumeyrolle}, namely a symplectic Clifford
group, $\mathscr{G}_S(V)$, covering the spinoplectic (or toroplectic)
group $Sp_2 (V)$, which is a non-trivial double cover of the
symplectic group $Sp(2n,V)$, where $V$ is assumed to be a real space
and $dim (V) = 2n$.  There is also a metaplectic group, and higher
covering spinoplectic groups $Sp_q (V)$, $q > 2$.  There are
subtleties with the symplectic Clifford algebras which we will
overlook for the moment, since the common symplectic Clifford algebra
(Weyl algebra) is a polynomial algebra, so that some additional
structure must be added in order that the exponential map be defined.
There are a number of alternatives for this extra structure.
See~\cite{crumeyrolle} and also Section~\ref{sec:spinors} below.

We will describe the construction of a symplectic Clifford algebra in
the following section which is distinguished from the symplectic
Clifford algebras in~\cite{crumeyrolle} (although it may contain some
of those algebras) which is better adapted to our purposes and in
which the exponential map is well defined, without arbitrary
assumptions, although our description
will not be exhaustively complete.  Thus, our conjugation is by an
element of the semigroup of symplectic transformations, which for the
unitary sub-semigroup have a
covering structure of orbits of spinoplectic transformations acting
from the left only, and whose semigroup meaning is not qualified or
restricted in any way once identified (locally but uniquely) with
the appropriate spinoplectic covering structure.  As to the unitary
symplectic transformations, our (locally defined)
conjugation fixes a group covering structure which is global in
some sense, analogous to the orthogonal case, but full invertibility
in the sense of a full group of transformations of the symplectic family
of groups need not extend from
the unitary transformations to the full group of symplectic transformations.
There may be group covering structures to the symplectic group (or there
may not be -- we do not inquire further into this issue here.) We still associate
dynamics with the symplectic transformations, which we are able to
define in general only in semigroup form, and not with the spinoplectic
and other covering structures, which may or may not be full group
structures.  It is only for the unitary transformations, a proper subset of
the symplectic transformations, that we will have assurances of a full
and proper group structure (up to sets of measure zero). 
Topological obstructions exist because the geodesics generated by
the full symplectic Lie algebra include the orbits of hyperbolic generators,
for whom inverses are only locally and infinitesimally
uniquely defined, which is why macroscopically there are semigroups and
not full groups of symplectic transformations. 

We have a very strong motivation for working with
Clifford algebras and their representations: we know instantly that we
have a covering structure for which our constructions are well
defined and have some sense of global meaning.  Below, we
establish only that the unitary transformations form an invertable
subgroup structure within the semigroup of symplectic
transformations--we will view the unitary transformations as
forming the nucleus of larger families of transformations, and
in our immediate concerns it is only the macroscopic invertability
in fact of the unitary transformations which is necessary for
our present constructions to be well defined.  One of our
implicit lessons is that the symplectic Clifford constructions
do {\em not} generally result in analogues to
equation(\ref{eq:orthrot}) and equation (\ref{eq:spinrot}),
and that the invertability of the unitary transformations
within the symplectic Clifford algebras may be topologically
dependent on their invertability in the orthogonal Clifford
algebra.  There is an important result in the next
section that the general
symplectic transformations are topologically obstructed from
invertability and are inherently semigroup in nature, with
the physical consequence that any sufficiently general
hamiltonian dynamics is not macroscopically invertable, i.e., for
general hamiltonian dynamics, the transition from equation
(~\ref{eq:orthrot}) to equation (~\ref{eq:spinrot}) must be
regarded as strictly local (infinitesimal).

\section{Spinors and Clifford algebras}\label{sec:spinors}

This brings us to an interesting juncture, which we illustrate with
the simple case of the phase space over a single pair of canonical
variables, which we will call $p$ and $q$.  Our phase space, which we
denote $T^\times \R$, has a basis $e_p ,e_q$.  If we perform an
analytic continuation of functions on $T^\times \R$, we will get a
space of functions over an ``analytically continued'' phase space
upon which we construct both an orthogonal and a complex
symplectic structure (which we take in its real or symplectic
form), and which we may label the complex extension of the underlying
phase space $\widetilde{T^\times \R} \equiv
T^\times \R \oplus i \circ T^\times \R$.  It is no great assumption
to regard both $T^\times \R$ and  $\widetilde{T^\times \R}$ as inner
product spaces (with alternative bilinear forms defining alternative
products.) Below, we define a
basis for $\widetilde{T^\times \R}$ which is simultaneously orthogonal and
symplectic (compatible with both orthogonal and symplectic forms in
their standard form), just as $e_q$ and $e_p$ provide such a basis for
${T^\times \R}$.  That basis is closely related to the
creation and destruction operators or $\pm$ the unit imaginary times a
creation or destruction operator (borrowing directly from
\cite{crumeyrolle}, page 247):
\begin{equation}
\textrm{Define}\;\; a= \frac{(e_q+ie_p)}{\sqrt{2}} \;\;
\textrm{and} \;\; a^\dag = \frac{(e_q-ie_p)}{\sqrt{2}}
\label{eq:crumops}
\end{equation}
so that the basis we choose for $\widetilde{T^\times \R}$ is
\bea
\epsilon_1=& ia\sqrt{2}   \qquad \epsilon_{1\ast } = a^\dag \sqrt{2}\nonumber \\
\epsilon_2=& a\sqrt{2} \qquad \epsilon_{2\ast}= ia^\dag \sqrt{2}\quad .
\label{eq:sobasis}
\eea
This is essentially just an orthonormal analogue of the creation and destruction
operators, stated in phase space rather than on our function space
representations.  Because they can identified with  particular values
of canonical
position and canonical momentum, 
$a$ and
$a^\dag$ obey the familiar commutation relations of the creation and
destruction
operators ($A$ and $A^\dag$ of~\cite{II}).  In phase space, we have
the Poisson bracket for $a$ and $a^\dag$ corresponding to the
commutator of $A$ and $A^\dag$ on our function spaces; this is
analogous to the Dirac canonical quantization $\{ p,q\}
\longrightarrow \frac{1}{i\hslash} [P,Q]$, except with analogues to
the creation and destruction operators on phase space, there is no
$\hslash$ associated with the canonical quantization, suggesting
something of a more fundamental geometric nature at work. See also the
Appendix.

This means we have broken $\widetilde{T^\times \R}$ down
into transverse hyperbolic spaces with bases $\{ \epsilon_\alpha\}$ and
$\{\epsilon_{\alpha \ast}\}$,$\alpha =1,2$, respectively.  These
satisfy the
relation
\bea
(\epsilon_\alpha ,\epsilon_\beta)=& (\epsilon_{\alpha\ast}
    ,\epsilon_{\beta\ast} )=0 \qquad (\epsilon_\alpha,
        \epsilon_{\beta\ast} ) = \delta_{\alpha\beta}
            \nonumber \\
\omega (\epsilon_\alpha ,\epsilon_\beta)=& \omega (\epsilon_{\alpha\ast}
    ,\epsilon_{\beta\ast} )=0 \qquad \omega (\epsilon_\alpha,
        \epsilon_{\beta\ast} ) = i\delta_{\alpha\beta}
\label{eq:spotests}
\eea
where $(\cdot , \cdot )$ is the symmetric form (e.g., associated with
the $\{ \cdot , \cdot \}_+$ symmetric bracket on phase space and
anticommutator on our representation function spaces, and with the
symmetric scalar product), and
where $\omega$ is the skew symmetric form (e.g., associated with the
familiar $\{ \cdot , \cdot \}_-$ Poisson bracket on phase space and
the commutator on our representation function spaces, and with the
skew symmetric scalar product).
Geometrically, $\epsilon_1 \perp \epsilon_2$ amd $\epsilon_{\ast 1}
\perp \epsilon_{\ast 2}$, because of the orthogonality of real and pure
imaginary components.  The other relations are straightforward.  The
$\epsilon_\alpha$ and $\epsilon_{\alpha\ast}$ thus form a real Witt basis
for the metric of $\widetilde{T^\times \R}$, the complexification of
the phase space $T^\times \R$.  Generalization to higher dimensions is
straightforward.

Note that to implement the constructions of~\cite{II}, it is necessary to
use the {\em commutative real algebra} $\C (1,i)$ for the ring of
scalars of our Clifford algebras, rather than the {\em field} $\C$,
for the reasons given in~\cite{II}.  We will also adopt the notion of
involution given there for our adjoint transformations.  This is
what~\cite{porteous} calls a $L^\alpha$ Clifford algebra, $L$ being a
commutative algebra and $\alpha$ being an involution.

Of related importance to us is the notion of ``correlation''.   The
scalar product $\langle \psi^{\textrm{out}} \vert \phi^{\textrm{in}}
\rangle$ establishes the correlation between the prepared state
$\phi^{\textrm{in}}$ and the observed effect $\psi^{\textrm{out}}$.
The matrix elements of quantum theory are representations of
correlations of this sort.  With our
multicomponent spin vectors there are more exotic correlations which
it is possible to calculate.  If on our space of states there is a symmetric
(orthogonal) form
\begin{equation}
Q=\begin{pmatrix} 0 & \II \\ \II & 0 \end{pmatrix} \quad ,
\nonumber
\end{equation}
where $\II$ is the appropriate unit operator, and if there is a symplectic form
\begin{equation}
F=\begin{pmatrix} 0 & \II \\ -\II & 0 \end{pmatrix}
\nonumber
\end{equation}
in addition to the scalar product
$\langle \psi \vert \phi \rangle$ we can form the symmetric (bosonic)
correlation $\langle Q\psi \vert \phi\rangle$ and the skew (fermionic)
correlation $\langle F\psi \vert \phi\rangle$~\cite{I}.  The unit
imaginary
``$i$'' is associated with a skew symmetric form as well, providing a
further source of skew correlation.  The significance of
the unit imaginary ``$i$'' is that it is associated in our
constructions with a complex hyperbolic (e.g., Lobachevsky) geometry,
rather than some real hyperbolic structure--both are associated with a
symplectic form~\cite{II}.  The unit imaginary is a ``correlation
map''~\cite{porteous}.

The point of this preliminary bit of algebra is that in our construction in
part two of this series~\cite{II}, we used the creation and destruction
operators to form polynomials acting as generators of (geodesic)
transformations.  Their role as vectors there is fully equivalent to their use
within this paper hereinabove: {\em this}
$a$, $ia$, $a^\dag$ and $ia^\dag$ are in nearly all regards identifiable
with {\em that} $A$ and $A^\dag$, etc., being an orthonormal (real Witt) basis
associated with the later by taking special values of $p$ and $q$ in the phase
space.  In particular, they allow us to define an  orthonormal basis according to
the hyperbolic scalar product defined by the symplectic form.
Here they provide a basis for the base space of a
symplectic Clifford algebra and at the same time they can provide a basis for
the base space of an orthogonal Clifford algebra for the
same space (the complexification of phase space).  Both
orthogonal and symplectic Clifford algebras of a single phase space
exist simultaneously in this basis!  In~\cite{II}, the creation and
destruction operators were used to build the generators of
infinitesimal translations (vectors) on the space used for the representation
of hamiltonian dynamics.  The roles of the $a$ and $a^\dag$ and the $A$ and
$A^\dag$ are comparable on their respective spaces.  In particular, $A$
and $A^\dag$ may be substituted for $a$ and $a^\dag$ in equation~\ref{eq:sobasis}
and the relation equation~\ref{eq:spotests} still holds: creation and annihilation 
operators provide a representation of the real Witt basis of the complexification
of phase space.

The orthogonal Clifford algebras can be formally defined using the tensor algebra of a
space.  Thus, given a (real) space $E$ space with a symmetric bilinear (quadratic) form $Q$
defined on it, the orthogonal Clifford algebra $Cl_O (E)$ is defined as
the quotient of the tensor algebra $\times (E)$ by the two sided
ideal ${\mathscr {N}} (Q)$ generated by elements of the
form~\cite{crumeyrolle}, p. 37:
\begin{equation}
x \times x - Q(x) \; , \quad x\in E \subset \times (E) \; .
\label{eq:cl0}
\end{equation}
Due to invocation of the tensor algebra, existence is a fairly trivial
issue.  The orthogonal Clifford algebras contain the orthogonal
Clifford group, ${\mathscr G}_O$, which contains the familiar pin,
spin and orthogonal
groups as subgroups.  The orthogonal Clifford algebras are associated with the
representations of fermions.  (The familiar Dirac algebra is the even
subalgebra of the orthogonal Clifford algebra for Minkowski space with
signature $(+,-,-,-),$~\cite{hestenes0},~\cite{hestenes}, pp 67, 75.)

Similarly, for $E$ an $n$-dimensional real vector space, and $F$ an
antisymmetric bilinear form, we define the common symplectic Clifford algebra
$Cl_S (E)$ by the quotient of the tensor algebra $\times (E)$ by the
two-sided ideal ${\mathscr{N}} (F)$ generated by the elements~\cite{crumeyrolle}, p. 233,
\begin{equation}
x\times y - y\times x - F(x,y) \; , \quad x,y \in E \; .
\label{eq:cls}
\end{equation}
The common symplectic Clifford algebras (Weyl algebras) are
essentially polynomial algebras, for which the exponential map is not
closed, and
comprise subsets of a number of other
symplectic Clifford algebras.  Of particular interest to us are the
formal symplectic Clifford
algebras over $K((h))$ containing a symplectic Clifford
group, ${\mathscr G}_S$, which contains the toroplectic (metaplectic),
spinoplectic and familiar symplectic groups all as subgroups.  We
identify the ring $K$ as the commutative real algebra $\C (1,i)$, and
$h$ is Planck's constant.  (See~\cite{crumeyrolle} for details.)
The symplectic Clifford algebras are associated with the
representation of bosons.

If $Cl_O (E) = \times (E) / {\mathscr{N}} (Q)$ and $Cl_S (E) = \times (E)
/ {\mathscr{N}} (F)$, are both defined for a space $E$,
there are obvious generalizations, such as, in particular,
\begin{equation}
Cl_U (E) = \times (E) / \left \{ {\mathscr{N}} (Q) \cup {\mathscr{N}} (F)
    \right\} \quad .
\label{eq:clu}
\end{equation}
It is straightforward that
\begin{equation}
Cl_U (E) = Cl_O (E) \cap Cl_S (E) \ne \emptyset
\label{eq:clualt}
\end{equation}
$Cl_U (E)$ contains the space $E$ and $\II_E$, so is non-trivial.  We will
call it the unitary Clifford algebra, and note that its existence
depends on the space $E$ having a real Witt basis (such as phase space
or its complexification) and both symmetric and antisymmetric forms
(such as phase space or its complexification).  At this point, we must
regard it as a set in the tensor algebra having algebraic properties,
although not necessarily a fully endowed ``algebra'', and in
particular the exponential map is not closed on it
(since this is the case for the Weyl algebras).

The orthogonal Clifford algebras have an exponential map which is
complete as a by product of their ring of scalars being the reals,
effectively using the same norm topology as $\R^n$.
See~\cite{crumeyrolle}.  Choosing this same separating norm topology
for the completion of sequences formed of elements from $Cl_U (E)$,
we can form what we will temporarily call the
orthogonal completion (o-completion) of the unitary Clifford algebra,
which we denote $Cl_{U-O}(E)$.  It follows that $Cl_{U-O}(E)\subset Cl_{O}(E)$.

The common symplectic Clifford algebras (Weyl algebras) do not contain
Lie groups since they are basically polynomial algebras and
the exponential map is not complete in them.  Some
form of completion may be imposed on them in order to obtain an
augmented symplectic Clifford algebra in which the exponential map
converges.  A linear topological space is an algebra plus a scalar
product, so it is no great additional assumption
if we treat $Cl_{U-O} (E)$ as a topological
vector space complete in the o-topology as indicated above.
We may freely regard our base spaces--phase space and its
complexification using the commutative ring
$\C (1,i)$--as scalar product spaces.  Since the
space $Cl_{U-O} (E) = Cl_{U-O} (\widetilde{T^\times \R^n})$ is now a
linear topological space, it
has a neighborhood of $0$ of sets complete complete in $Cl_{U-O}
(\widetilde{T^\times \R^n})$ in any finer topology than the
o-topology previously
chosen for the completion of $Cl_S (\widetilde{T^\times \R^n})
\cap  Cl_U  (\widetilde{T^\times \R^n})$, and
we obtain thereby a complete linear topological space in the
finer topology.

Thereby, the Kobayashi
semidistance on the complex hyperbolic space $\widetilde{T^\times
\R^n}$, $n\ge 2$, provides a (effectively seminorm or weak) topology
for $Cl_U  (\widetilde{T^\times
\R^n})$~\cite{lang}.  The notion of geodesic is well defined in this
topology~\cite{lang}, which we will call the s-topology.
Convergence of symplectic (=dynamical) transformations in
semigroups is thus a sufficient condition for us to talk about
bosons, e.g., a dynamical  arrow of time is a sufficient condition for us to
talk about bosonic fields continuously evolving in this construction.

By these standard theorems, $ Cl_{U-O}
(\widetilde{T^\times \R^n})$  will also be complete in the finer seminorm
s-topology~\cite{schaefer}.  Local convexity, and so on, easily follow from
this.  In the sequel, we will work with the algebraic set
$Cl_U  (\widetilde{T^\times \R^n})$ as a completed linear
topological space $Cl_{U-O} (\widetilde{T^\times
\R^n})$, with alternative normed o-topology and seminorm
s-topology completions\cite{note2}. We will distinguish the individual
completions, as necessary, by indicating the form of completion thus:
$ Cl_{U-O} (\widetilde{T^\times \R^n})$ and $ Cl_{U-S}
(\widetilde{T^\times \R^n})$.

Putting matters slightly differently, we can use the real Witt basis defined
above as the basis for a unitary Clifford algebra, the intersection of
the orthogonal and symplectic Clifford algebras of the
complexification of phase space:
\begin{equation}
Cl_U (\widetilde{T^\times \R^n}) = Cl_O (
    \widetilde{T^\times \R^n}) \cap
        Cl_S (\widetilde{T^\times \R^n}) \quad n\ge 2 \quad.
\label{eq:cluphasespace}
\end{equation}
From the perspective of o-topology associated with the orthogonal
form, we may identify $Cl_{U-O} (\widetilde{T^\times \R^n})$ with
representations of fermionic particles.  From the perspective of
s-topology
associated with the symplectic form, we may identify this same
$Cl_{U-S} (\widetilde{T^\times \R^n})$ with the representation of
correlated hamiltonian dynamics of bosonic
fields.  The complexification of phase space, $\widetilde{T^\times
\R^n}$, $n\ge 2$, is used for the representation of correlated
dynamics over $T^\times \R^n$.  We are dealing with the algebraic
treatment of correlated dynamics from alternative perspectives by
using the vehicle of Clifford algebras and their representations.

According to this prescription, all the spaces in our Gel'fand
triplets of spaces in the
Gadella diagrams are built by using alternative topological
completions of Clifford algebra representations.  The
Kobayashi semidistance adapted to provide a seminorm above does not
produce a countable family of seminorms, such as involved in the
construction of our representation spaces in~\cite{I}, although the
topological completions obtained through its use are locally convex
spaces with a nuclear part of our unitary Clifford algebras.
Note that if phase space is taken as a scalar product space, it is
then straightforward to define a rigged Hilbert space over
the locally convex nuclear space
$\overline{Cl}_{U-S} (\widetilde{T \R^n})$ as follows,:
\begin{equation}
\overline{Cl}_{U-S} (\widetilde{T \R^n})\subset
  \overline{Cl}_{U-O} (\widetilde{T \R^n}) \cong
   \overline{Cl}^\times_{U-O}  (\widetilde{T^\times \R^n}) \subset
    \overline{Cl}^\times_{U-S} (\widetilde{T^\times \R^n}) \quad ,
\end{equation}
where the over bar relates to the nuclear locally convex parts only.

\subsection{A tentative physical view}\label{sec:tentative}

In the preceding section, we indicated how the difference between
fermions and bosons in one of choice of topological completion, with
the bosons being associated with a finer topology, i.e., a topology
which separates more points than does the topology used to construct
the fermions.  There is a natural way of translating this into
physics.  When the Compton and de Broglie wavelengths are comparable,
the physical phenomena are intrinsically quantum and the wave nature
is in evidence.  It seems natural to associate
the Compton wavelength with our finer topology and bosons of
$CL_{U-S}(\widetilde{T^\times \R^n})$.) We would infer that in the coarser
topology of $Cl_{U-O} (\widetilde{T \R^n})$, the Compton wavelength may be
smaller than the de Broglie wavelength and the particle nature is in
evidence.  This is of course reasoning by analogy, and not to be taken
as law, but can explain such phenomena as why one never observes a
free quark, for instance: there is no such thing (yet) as a
non-relativistic free quark, just as there is no such thing as a
non-relativistic photon.
Thus, it should not be taken as canon law that topology has no
observable consequences, in the sense that a choice of topology may
reflect an assignment of relative mass and energy scales, etc., to the
phenomena undergoing mathematical description.  One has to adapt the
mode of description to the phenomena being studied, and this may mean
choosing a particular topology (perspective) for the representation.
(Possible distinctions between bosonic and fermionic representations
are given in~\cite{hanany}, and we would conjecture this has observational
consequences.)

\subsection{Why focus on the unitary Clifford  algebra?}\label{sec:focus}

Our ultimate goal is to define a set of structures in which every
space in the associated Gadella diagrams is a spin space, since we
have already seen in installment two~\cite{II} the presence of
multicomponent vectors (which in fact satisfy technical requirements
for being spinors).  The unitary Clifford algebras are significant
because they contain the
relevant unitary group (or semigroup) as transformations groups, but
they play a special role for us because they contain the relevant
special unitary
group (or semigroup).  The unitary groups preserve the Witt bases
which are the foundation of our construction.  Because we have complex
{\em simple} Lie groups, and not merely semisimple Lie groups, we are
assured of unique spin structures~\cite{lawson}.

We will focus on the special unitary groups, rather than the full unitary
groups, because $SU(N)$ is simply
connected, and spin manifolds are manifolds with simply connected
structure groups, relevant to having well defined spinor bundles
associated to the $SU(N)$ generated dynamical flow structure which we
will use to set up a gauge theory in the following
section.  Also,
$SU(N)$ has a bi-invariant Riemannian (symmetric) metric, and this
metric is identifiable with harmonic forms, meaning that the group
(and associated flows enerated by it) will have a well defined
harmonic structure, with kernels and propagators, etc., well defined
(initially in the o-topology).  Likewise, one parameter subgroups are
geodesic.  $Cl_{U-O} (\widetilde{T^\times \R^n})$ (and its function space
representations) is thereby a very
well behaved linear space, and associated to it are well defined flows and
harmonic structure on $\widetilde{T^\times \R^n}$.  We thus have
great confidence that  our mathematics is well defined when working
with special unitary groups,
although it may be possible that such well-behavedness could be extended
to the full unitary groups.

The spin geometry for $Cl_O(E)$ is a specialized branch of Riemannian
geometry~\cite{lawson}.  For $Cl_U(\widetilde{T^\times \R^n})$, when
the unitary Clifford
algebra is completed in the category of topological linear spaces
using a seminorm topology, the real Witt basis gives us a vehicle
to obtain a well defined spin
geometry for the simply connected (locally convex, nuclear) completion
$\overline{Cl}_{U-S} (\widetilde{T^\times \R^n})$ as follows.  The
well defined spin structure on $Cl_{U-O} (\widetilde{T^\times \R^n})$
derives from the spin structure on  $Cl_{O} (\widetilde{T^\times
  \R^n})$.  The first and second Stieffel-Whitney classes are trivial
on $Cl_{U-O} (\widetilde{T^\times \R^n})$, and are homotopy invariants
(characteristic classes).  (For proper homotopy theory, we must use
groups and not semigroups.  See Section~\ref{sec:spinbundles} below as
to the semigroup $SU(N)_\pm$ having no obstructions on any set
of positive measure to extapolation to a full group structure.)
It follows that this spin structure survives
the change to a finer topology so that $\overline{Cl}_{U-S}
(\widetilde{T^\times \R^n})$ also has a well defined spin structure
(and harmonic structure, etc.).

Because $exp$ maps
dense sets to dense sets (topological notions!), even for
non-compact (e.g., hyperbolic)
generators, the $\overline{Cl}_{U-S}(\widetilde{T^\times \R^n})$
can serve as a nucleus for a covering space of the common symplectic
 Clifford algebra or Weyl algebra
$Cl_S (\widetilde{T^\times \R^n})$.  By this
device, we obtain a complete nuclear (possibly locally convex)
linear topological space, the nuclear  symplectic Clifford algebra
$\overline{Cl}_S (\widetilde{T^\times \R^n})$ for which the exponential
map is complete.  This suggests that there is in some qualified sense
a well defined spin geometry on all or at least parts of
our nuclear symplectic Clifford algebra,
$\overline{Cl}_S (\widetilde{T^\times \R^n})$, and that the exponential map of
the Lie algebra of the symplectic semigroups $Sp(2n,\R )^\C_\pm$
is holomorphic with respect to our seminorm completion,
and as extended in this seminorm topology $\overline{Cl}_S(\widetilde{T^\times
  \R^n})$ is also a Clifford algebra containing the common symplectic
Clifford algebra (Weyl algebra) as a sub-algabra.  (Spin geometry as
currently understood is dependent on a simple connected structure group--the
qualified sense of a well defined spin geometry as referred to preceding
may require an extension of spin geometry as currently accepted.  The topological
obstructions that leave us with semigroups rather than groups do not
obstruct the exponential map.  The nuclear
symplectic Clifford algebras are locally path connected, but may not be simply
connected. We are not
concerned with these and related issues herein, but there are obvious issues
that should be explored.)

As a noteworthy aside, since our Clifford algebras
above are completed in the category of linear topological spaces, they
also possess their own Clifford algebras.  Thus, we have the
possibility of constructing towers of algebras, and these have
properties of interest also.~\cite{coxeter}  In fact, as to the connected part,
since we have a
nuclear locally convex topology, these Clifford algebras may serve as
the base space of a Gel'fand triplet, e.g., the $\F$ of the RHS
$\rhs$: for the abstract base space
$\F_{\mathfrak{sp} (2N,|R )^\C}$ of the Gel'fand triplets of
~\cite{I} and~\cite{II}, we could
take $\overline{Cl}_{U-S} (\widetilde{T^\times \R^n})$,
(or possibly $\overline{Cl}_{S} (\widetilde{T^\times \R^n})$).
There are thus towers of Gel'fand triplets over the
orthogonal and symplectic unitary Clifford algebras. The unitary Clifford
algebras (both o-completion and s-completion) provide finite dimensional
representation spaces of the (e.g., compact) unitary group and also
provide a tower of (finite dimensional) Hilbert spaces.  These
infinite tower structures bring to mind the scale of Hilbert spaces,
$ \cdots \HH_n \subset \HH_{n+1} \subset \cdots \subset \HH_\infty$,
of ~\cite{genfun4}.  This scale may be relaxed somewhat, and the
mathematics is still sufficient to do interesting post-Hilbert space
physics~\cite{antoine}, and indeed the Schwartz space $\CS$ may be
obtained from taking the intersection of the spaces $\HH_n = D(\R^n
)$.  Our constructions seem to operate in the convergence of a lot of
well defined mathematics with post-Hilbert space physics.

As to the unitary Clifford algebras, it is reasonably straightforward
to obtain fiber bundles with unitary structure groups, much in the
manner typical for frame bundles obtained from the base space and
orthogonal transformations of an orthogonal Clifford
algebra.  In similarly straighforward and well known manner, one may
obtain bosonic and fermionic principal fiber bundles with special unitary
groups as the structure group.  (We will
discuss these and their physical relevance below.)  There is also a
suggestion of a type of ``dynamical principal fiber bundle'' with
semigroups of symplectic transforms as structure (semi-)group~\cite{note5}.  We
will provide a description of the spinor bundle structures of
immediate relevance in Section~\ref{sec:spinstructure}.

Note that  in general the Hamiltonian does not commute with the full
symplectic Lie algebra, so that energy is not a constant of all
possible dynamical evolutions (i.e., it is possible to represent
open systems), and the energy
eigenstates do {\em not} provide an irrep of the group--typical for
spinor representations of groups.  In order to include complex
spectra in a mathematically well defined formalism, we have been
led by {\em mathematical necessity} to representations
which are neither unitary (they are ``dynamical'', or more general)
nor irreducible (they are spin)!

There are also other possible implications of defining the unitary and
extended symplectic Clifford algebras as we have.  The orthogonal
Clifford algebras are associated with commutative
geometry~\cite{connes,garcia-bondia}.
Because the symplectic Clifford algebras associated with the skew
symmetric symplectic form rather than the symmetric orthogonal form,
it is possible (and we will so conjecture) that our s-topology
completion of the unitary Clifford algebra and the extended Clifford
algebra obtained from it are associated with some type of noncommutative geometry.
The Clifford algebras (all types) are $\mathbb{Z}_2$ graded and thus are
superalgebras; when topologically completed as spaces they are
superspaces as well.  Although beyond the scope of these present
inquiries, we will conjecture that most of the machinery of
noncommutative geometry, superalgebras and superspaces, Hopf
algebras, etc., (but not SUSY) is fairly close to hand even though not
presently
revealed.  If these speculations are true, the unitary Clifford
algebras possess both commutative and noncommutative geometric
structures, depending on the choice of perspective (choice of bilinear
form and topological completion).  Another instance of
the unitary Clifford algebras seeming
to be a regime in which a lot of mathematical machinery is
exceptionally well behaved, connecting a lot of disparate
methodologies by having them defined over the same sets.

The spinor discussions in Section~\ref{sec:necessity} refer to
invertibility of what are nominally semigroups in the context of a
covering structure which is spin.  In the context of the orthogonal
transformations, the well known spin groups provide the simply
connected covering structure for obtaining equation (\ref{eq:spinrot})
from equation (\ref{eq:orthrot}), even in the case of semigroups of
orthogonal transformations--simple connectedness is the key.  The
spinoplectic {\em groups} provide the analogous simply connected
covering structure for the semigroups of symplectic
transformations~\cite{crumeyrolle}.  Thus, even though our use of a
seminorm topology for the complex symplectic Clifford algebras
formally results in semigroups of transformations, there are no
obstructions on sets of positive measure to our extrapolating simply
connected sub-semigroups of the unitary semigroup
into full group structures--we can convert
special unitary sub-semigroups such as
$SU(4)_\pm$ semigroups into full groups.  These special unitary
(effective) groups may be thought of as subgroups of symplectic and
spinoplectic {\em groups} (in yet another topology!)  We therefore conclude
that the unitary sub-semigroups of our extended symplectic Clifford
algebras effectively provide a spin representation of the special
unitary group (and possibly the unitary group) within both s-topology and
o-topology completions of the algebraic set $Cl_U (\widetilde{T^\times
  R^n})$.

\subsection{Physical consequences}\label{sec:consequences}

We will ultimately adopt a gauge field interpretation for these
constructions, and this approach has some interesting physical
consequences.  Thus, dynamics should be mediated by bosons, as
represented by the spinors which in turn belong to
a representation of some form of a symplectic Clifford algebra.  On
the other hand, bulk matter
(fermions) should be represented by spinors representing an orthogonal
Clifford algebra.  Because there are more generators for $Sp(2n,\R )$
than for $U(n)$, there is the formal possibility there could be bosons
(intermediaries for dynamical forces) with no direct coupling to bulk
matter properties. Likewise, we infer there are
aspects of bulk matter not immediately associated with dynamics--i.e..,
apparently
the gravitational force does not depend on the kind of bulk matter,
but on the quantity of mass only.  The true quantum
geometrodynamics is contained only in the intersection of geometry and
dynamics, the unitary Clifford algebra.  We are working in a formal
system in which there is only a limited overlap in which we can
concurrently talk about all of the issues which are important to
us.  We are constrained to two separate perspectives, dynamics or
geometry, which do not completely overlap.  We must
choose one or the other perspective exclusively when we choose to speak
carefully, since there is no mathematically respectable way
of speaking from both perspectives at once.  There is,
however, a domain of strong correspondences in which a single
structure (an abstract spinor) may have alternative fermionic (even
dimensional skew symmetric matrix) and bosonic (odd dimensional
symmetric matrix) representations.  A
reminder that the representative is not necessarily the thing itself,
and that isomorphism may not mean equivalence in all senses.

Our abstract unitary Clifford algebra possesses both symmetric and
skew symmetric representations, corresponding to the bosonic and
fermionic perspectives.  In our carefully constructed mathematical
structures, the notions of boson and fermion correspond to field (e.g., wave) and
particle perspectives, respectively, but they no longer retain all of
their traditional meaning (which we would argue arises from working in
a mathematical formalism incompatible with resonances).  The presence of
composite bosons made up of fermions and recent developments in atom optics
give examples of why the topological wave--particle assignments should 
probably not be taken in any absolute sense. There is nothing which
prohibits bosonic and fermionic representations of the same thing, at
least on occasion.

\section{Spin Bundle Structures}\label{sec:spinstructure}

\subsection{Non-trivial dynamics}\label{sec:nontrivial}

At this juncture, let us recapitulate the road to the mathematically
well-defined covering structure for our hamiltonian
quantum field theory, incorporating resonances and other
coherence structures, and which is treated as a form of
dynamical system.  We based this on a probability rather than point
particle localization description of a dynamical system in phase space
(e.g., over canonical coordinates).  We tacitly assume that there is a
lot of freedom in the dynamical system and also that
we have some uncertainty in
our specification of the initial state.  A probabilistic field theory
is the result.  We added the
notion of correlation maps (injective embedings in the dual--which are
related mathematically to conjugation and associated notions of
coadjoint orbits, and related dynamically to momentum maps).  For
real symplectic (=dynamical) correlations of a simple two component
system, we
found that the system was either stationary or exponentially decays to
some equilibrium configuration.  There are more elaborate
correlations--complex symplectic (=dynamical)
correlations--which make more complex
behavior possible, with complex spectra and possible oscillatory time
evolution or damped oscillations occuring in the dynamical time
evolution of the probability amplitudes.  (E.g., such familiar
phenomena as diffraction patterns. etc., are evidence of {\em complex}
dynamical correlations.)

If we conjecture well behaved algebraic and topological properties
for the constructions of the preceding installments,
with proper algebras and topological linear spaces, we
are led to multicomponent representations, our function spaces
representing the probability amplitudes are $L^2$ spaces, and this
coupled with the
complex spectra forces us into a variant of the rigged Hilbert space
formalism such as was outlined in ~\cite{I} and ~\cite{II}.  The lesson of this
fourth installment is that these multicomponent vectors are indeed
spinors, and we identified the special roles played by the
unitary Clifford algebras in providing a very well defined
mathematical structure which forms a nucleus which may be enlarged to
provide covering structures so that all of the relevant dynamics and
geometry may be at least reasonably well defined mathematically. We
understand that our well behaved spinor structures are well defined
only in a special choice of basis, a real Witt basis.

The classical function space realizations
of our abstract RHS, $\rhs$, was shown by Gadella to belong to the
intersection of the Schwartz space ($\CS$) and the spaces of Hardy
class functions from above and below ($\HH^2_\pm$)~\cite{gadella}.
There are Clifford analogues of $\CS$ and
$\HH^2_\pm$~\cite{brackx,delange,gilbert}, and so the
function space realizations will be well defined if the abstract
spaces are also well defined.  (Recall the Gadella diagrams of
installment one~\cite{I}, and references therein.)

Gadella's use of van Winter's theorem~\cite{gadella} still provides the
necessary and sufficient conditions for ``analytic continuation''.
Just as in the work of Bohm~\cite{bohm1,bohm2}, there will be contours at
infinity in integrals.  We have performed our construction in such a
way as to define and then preserve under dynamical transformation
the complex hyperbolic (Lobachevbsky) geometry of
the tangent space to our space(s) of states.
The functions spaces used are classic Schwartz and Hardy
spaces as to their components, and so the necessity proof of
Gadella-van Winter suffices even in the spinorial RHS paradigm.

We identified in~\cite{III} many possible linkages with the classical
treatments of dynamical systems, and in particular possible
relationships with various notions of complex systems, statistical
mechanics (and related thermodynamic ideas), fractals, etc.,
which seem compatible with the formalism,
but which emerged as a by-product of largely mathematical
considerations in our incorporation of correlation and associated
resonances into a classical probability description of hamiltonian
dynamics on phase  space.  The question arises then whether this is
another instance in
the long string of unreasonable effectiveness of mathematics in
physics described by Wigner many years ago, or if we have wandered off
the path somehow.  In the following subsection,
we will adapt this structure to exhibit principal bundle structures
associated to our constructions, and interpret this structure
as a gauge theory, setting the stage for calculations which make
predictions which will ultimately
tell us if this is a toy theory or has some relevance to the real
world.

\subsection{Special unitary spinor bundles}\label{sec:specialunitary}

The unitary groups are compact and locally path connected,
while the special unitary groups are simply connected, with geodesic
subgroups.  It might be supposed that because our sought after
spin structure is obtained from a seminorm topology, there is no
invertibility, notwithstanding that $SU(N)$ is compact and simply
connected. The inverses used in conjugation are in one sense
basically pullbacks along a single fiber to the nucleus (our base
space), and so are well defined individually, but in general may
be well defined only locally. The crucial issue then is the issue
of whether or not there is a spin structure which will make 
global identifications possible.
We will examine the existence of spin structures further below,
but all these structures (and, for present inquiries, especially
the spin structure) depend on both the existence of and
choice of a special basis, the real Witt basis, and we must suppose
that there would not be invertibility of any sort
in a general basis.  Given the extreme dependence on the choice of
basis, which is a typical feature of any spin construction,
 there is an interesting interaction between a weak (seminorm/semidistance)
topology (and associated semigroups), spin conjugation and momentum
maps worthy of much further inquiry than will be undertaken in this first
description.  Our spin conjugation is not an
``inner automorphism'', but is a momentum map, involving the dual,
reinforcing our choice of completion of the unitary Clifford
algebraic set as a linear topological space, with scalar product and
dual.  The existence of any principal bundle structure depends on the
triviality of the structures, or, equivalently, on the existence of
sections. Recall that with our spinors in the orthogonal
case a locally defined conjugation was equivalent to a spin
transformation that acted from the left only and which was
part of a globally defined structure. This globally definition should
survive change to a finer topology, as outlined in Section~\ref{sec:focus}.
This suggests the probability of extending bosonic transformations
outside of the unitary core discussed earlier and also below, but we will not tackle
that issue directly in this forum, and in the next subsection we indicate
grounds to believe such an effort should fail.

With respect to the special unitary semigroup orbits in
 $\F_{\mathfrak{su}(N)\pm}\subset \F_{\mathfrak{sp}(2N, \R )\pm}$,
notwithstanding the seminorm topology, there is no obstruction to
invertibility on any set of positive measure: we may regard the
entirety of $\F_{\mathfrak{su}(N)\pm}$ as simply connected
fiber liftings of a simply connected base space.
Conventionally, if given
a space $E$ with base space $B$ and and whose fiber $F$ is isomorphic
to group $G$, a principal fiber bundle structure associated to $E$ has
a global section (making both $P(E)$ and $E$ trivial).  This means
there is a continuous mapping
\begin{equation}
s:B \; \longrightarrow \; E
\label{eq:sectionmap2}
\end{equation}
which is invertible, i.e., there also exists a projection $\pi$ such
that
\begin{equation}
\pi s(x) = x \quad , \forall x\in B \;\; .
\label{eq:projectionmap2}
\end{equation}
This means, in effect, that $s=\pi^{-1}$.  See, e.g.,~\cite{nash}.  The
lifting $s=\pi^{-1}$ is in fact all we really have globally for the
full symplectic (semi-)group in the
present case.  The projection $(\pi )$ is not defined as a continuous
transform on $\F_{\mathfrak{sp}(2N,\R )^\C_\pm}$, due to topological
obstruction associated with our semidistance (weak or seminorm) topology.

Maximal compact subgroups are homotopy equivalent to the
Lie groups that contain them, e.g., $U(N)$ is homotopy equivalent to
$Sp(2N,\R )$.  However, homotopy is based on groups, and semigroups
won't do!  Thus, there are finite dimensional UIR's of $U(N)$, but
none of the noncompact $Sp(2N,\R )$, recalling
Wigner's definition of
noncompact groups.  This suggests, once again, that we should think
naturally of semigroups of symplectic transformations, and not of
groups--else, from this homotopy equivalence, one would expect
there to be finite dimensional representations of
$Sp(2N,\R )$ which are merely lifts of of finite dimensional $U(N)$
UIRs. (Similar statements could be made for other noncompact groups
containing compact subgroups.)

However, there is no obstruction to invertibility as to the special
unitary subgroup of the symplectic group itself.  We have simply connected
fibers and a simply connected base space: the lifting of the base space
are $1:1$ and onto the ``sections'', and so are isomorphisms and
invertible~\cite{note25}.
Thus, identifying (both abstractly and as to the related very well
behaved function space realizations) for instance,
\bea
B &\Longleftrightarrow& \left\{ |n_A\rangle\right\} \oplus
    \left\{ |n_B\rangle\right\} \nonumber \\
G&\Longleftrightarrow& {exp} \; \mathfrak{su}(N)^\C=SU(N)^\C
    \nonumber \\
F&\Longleftrightarrow& \textrm{span} \left\{\left( \theta\circ SU(N)^\C \right)
    \circ b \right\}  , \quad b\in B \nonumber \\
E&\Longleftrightarrow& \F_{\mathfrak{su}(N)^\C}
\label{eq:indentify}
\eea
where $\theta$ is the representation mapping $\theta : SU(N)^\C
\longrightarrow
\textrm{Aut} (E)$, and, in the first instance,
the representation {\em must} be framed in
terms of the creation and destruction operators,
i.e., the real Witt basis.  We here have chosen to represent $B$ by
using the simple
harmonic oscillator number states as a basis, e.g., the energy
representation.
$E$ locally has the structure $B\times F$ by construction, and we can
readily invert the represention homeomorphism to identify an element
of $SU(N)^\C$, so that we have $[\II\times\theta^{-1}] : B\times F
\longrightarrow B\times SU(N)^\C$, $[\II\times \theta^{-1}]\circ \left( x,
g(x)\right) \longmapsto \left( x,g \right)$, $x\in B$, $g\in SU(N)^\C$.
If in our candidate
for $P(E)$ we consider $s(x) \in SU(N)^\C$ and $g\in SU(N)^\C$, then
$gs(x)$
belongs to the fiber over $x\in B$.  $P(E)$ thus
has the global structure of a product between the base space $B$ and a
fiber $SU(N)^\C$.

Because as to the $SU(N)^\C_\pm$ sub-semigroup
structure we have no obstructions (on sets of positive measure)
to extrapolating the semigroup structure due to the weak Kobayashi
semidistance 
topology into a full group structure,  we have a candidate for a
dynamical homotopy group  for the base space.  Reiterating, there is
no such thing as a
homotopy based on semigroups (possible distributional measures, such
as Dirac measures confound the notions of continuity, analyticity,
etc.), and
the property of being spin is determined by characteristic classes,
which are homotopy invariants.  Our base is spin, our fibers are spin,
and so we may properly talk of spinor bundles, and principal spinor
bundles in particular, only as to the special unitary orbits within
the overall dynamical structure, and all our discussion must be based on a 
real Witt basis and its representations.  As indicated earlier, these may be
represented by either bosonic or fermionic spinors when we do take a
representation, depending on choice of bilinear form and associated
topology (``perspective''), and whether the representation is even or
odd dimensional.

\subsection{The full dynamical spinor ``bundles''}\label{sec:spinbundles}

If there were full groups
generally available for the full symplectic semigroups in this
construction, there would be no problem
thinking of the symplectic transformations of, for instance, the function
space of energy eigenfunctions of two free quantum harmonic oscillators,
which we will call $\CS_{\hh \pm}$, which produces a family of function spaces
we will call
$\CS_{\mathfrak{sp}(4,\R)^\C \pm}$.  This space may be
associated in some very loose (and as yet unspecified) sense
to a principal G-bundle, e.g., an $Sp(4,\R )^\C$-bundle with
$\CS_{\hh\pm}$ as base space.  However, the only initial suggestion of
invertibility is with the $U(1)$ sub-semigroups
which have local actions on $\F_{\mathfrak{sp}(4,\R)^\C \pm}$ (and the
related very well behaved spin-function spaces).
Invertibility generally does not extend to sub-semigroups larger than
$SU(N)^\C$, unless we can successfully implement globally defined spin
transformations, such as in the manner alluded to in the preceding subsection.  
If we were to extend the $SU(N)$ representation
nucleus into a representation of
the $Sp(2N,\R)^\C$ algebra, it seems as if there should exist
something like a spinor bundle (loosely, a vector bundle of spin type), but the
topological obstruction which keeps our semigroups from being
extrapolated into full dynamical groups probably also prevents formation of a
full spinor bundle structure, notwithstanding a proper spinor bundle
is contained (as a nucleus) somewhere within
this extended structure.  We do not have homotopy equivalence, and the
characteristic classes which define the property of being ``spin'' are
not preserved in arbitrary mathematical operations.  Properly, we have
special unitary spin bundles, with what we might call
``improper homotopies'' extending this
to a larger multicomponent ``symplectic bundle'' which is not spin in the
strictest sense, (just as the extended structure has no proper
principal bundle structure in any accepted sense). We can have a lifting
or a projection, but we cannot have both simultaneously, due to the
semigroup rather than group nature of the ``fibers''.

The dynamical liftings taken as a whole are not based upon
isomorphisms, as in the conventional treatment of principal
fiber bundles, since generally there is no invertibility to the lifting.  It
is precisely this lack of isomorphic
liftings of the ``paths along flows in phase space''--lifting of vector
fields composed of state vectors--which enables us to
convert two free oscillators
into a pair of coupled oscillators or vice versa.  If we identify this construction
with particle-fields, pair production or destruction it
is not a $1:1$ mapping, so the 
typical quantum resonance processes of pair production and annihilation
are not what we
would think of as an isomorphism either.  Dynamical pair production or
destruction is not $1:1$ in ${before}:{after}$, so in order to incorporate
such processes into our overall dynamical structure we have lost the
use of invertibility and thereby dynamical evolution is not an isomorphism 
in general.  The existence of pair production in nature confirms for us
that our non-invertible dynamical semigroup notions are sound physics. The
compact generators of $SU(N)$ take us from ``island of stability'' to
``island of stability'', while the noncompact generators of the full
symplectic semigroup,  which represents the full gamut of dynamics, 
take these ``islands of stability'' and make
resonances out of them, which will evolve dynamically towards another
``island of stability''.  Bifurcations are possible, in much the same
sense as that term is used in classical dynamical systems, except that
probability may flow along both paths of the bifurcation, e.g., there
may be pair production.

This lack of isomorphism in the ``fiber'' liftings raises questions
as to the extent to which we may reasonably think of the more general 
constructions over the full symplectic semigroups as principal fiber
bundles, which merely involve semigroups rather than groups for the
fiber of the ``principal bundle''. The lack, in general, of
invertibility in our structural semigroups, e.g., $Sp(4,\R
)^\C_\pm$, is reflected in the flow structure of the representation
space and provides a novel meaning to the term connection.  As in
the standard principal G-bundle construction, our Lie algebra
provides the ``connection''.  The path of $exp \,\g$ connects
``sections'' in both cases, e.g., $e^{-iHt}$ for $t\ge 0$ is the
semigroup which transports you from section (time slice) to section
(time slice) in generalized state space.  The Schr{\"o}dinger
equation is the equation for geodesic transport (parallel transport
in this case), giving us a constant of the motion: energy is
conserved.  There seems to be a clear sense of meaning here, and
clear mathematical analogies.  Thus, that  $e^{-iHt}$ is geodesic,
with a conserved quantity (energy) does not prevent time evolution
from being hyperbolic in appropriate cases, and yet time evolution
is ergodic with an equilibrium end-point (installment
three~\cite{III}, and recall the presence of fractals), though the
functions representing the system should be of bounded mean
oscillation (bounded analytic functions)~\cite{garnett}.  We would
therefore infer that in the present construction
the wave function for the universe as a whole
does not permit unbounded continuous creation, that the energy of
the universe has always been pretty much what it is now and will
remain pretty much the same in the future, although the universe may
continue to expand hyperbolically, to eventually become conformally
flat, etc.  (In Section~\ref{sec:breaking} we offer a speculative
interpretation of our bosonic spinors which offers an explanation of
how the Big Bang could conserve energy.) But, can we reasonably
treat this dynamical structure as some sort of fiber bundle, or at least 
what part of it may be so thought of?

These are physically appealing notions which which follow directly
from the mathematics, although of course they need substantial
elaboration to really make them respectable.  There are also
substantial physical interpretation issues to resolve, especially
those contrasting the gauge transformations in the o-topology versus
gauge transformations in the s-topology.  In equation
(~\ref{eq:decomp}), we pointed out in the introduction that one or
the other term vanishes because we will be expressing the operators
in terms of creation and destruction operators, and that one of the
two terms will vanish for either bose or fermi creation and
destruction operators due to the properties of those operators in
that perspective.  Yet, mathematically~\cite{narici} and physically
we should be amazed if a ``mere choice of topology'' has any
profound effect on the scalar product used to calculate the
expectation of any physical observables--it is not possible
physically to conduct a Cauchy sequence of experimental
measurements, so whether we use a bose or fermi realization of the
$SU(N)$ Lie algebra and Lie group should not effect the outcome of
the computation of this scalar product.  This means that, i.e., that
the probability of observation should be a topological invariant for
aspects of quantum dynamics associated with stability (like the
energy levels of atoms or the mass spectrum of the elementary
particles) and independent of bose--fermi notions, which are
topologically dependent ideas.

Of course, there is more than topology separating the larger
symplectic semigroup and orthogonal semigroup outside of their
coincidence--or overlap--on the unitary group. There is nothing like
any topological invariant there!  We have shown that in installment
two\cite{II} that the notion of resonance is associated with
dynamics and symplectic geometry and foreign to orthogonal geometric
notions, and hence we find differing energy eigenvalues depending on
topological choice when working outside of the overlap of the orthogonal and
symplectic semigroups of transformations--since on this domain
orthogonal geometry and its associated topology has nothing
whatsoever to do with dynamics! This suggests a clear divergence
from frame bundle notions at the very least. We will discuss 
homotopy notions (which are the underlying notions in conventional 
treatments of gauge theory) more fully in the sequel, but clearly algebraic
topology is different insignificant ways
when done in strong and weak topologies.

Note also that, since there can only
be a symmetric or skew-symmetric form on a space, the Grassmann
algebras (upon which the notions of supersymmetry are based), cannot
have any topological notions related to forms defined on them. Since
the notion of convergence of the exponential map is topologically
dependent, there can be no such thing as a unitary gauge group
associated with any Grassmann algebra, which further suggests that
notions of supersymmetry and notions of gauge group are
mathematically incompatible to the extent that supersymmetry notions
are identified with Grassmannian notions.

In addition, orthogonal gauge transformations are frequently regarded
as ``passive'', e.g., as simple changes of frame.  This, for instance,
is a common interpretation of a $U(1)$ gauge transformation in
electromagnetism, and the associated effect on the vector potential.
In installment two~\cite{II}, care was used to make possible
topologically transitive symplectic (=dynamical) transformations, and
as to the unitary transformations in the s-topology we are not
talking about ``passive'' frame transformations.  This is ``active''
dynamics
being represented in the s-topology on the dynamical gauge bundle,
notwithstanding we are talking of transforming one ``island of
stability'' into another.  This should be born in mind when reading
other parts of this Section~\ref{sec:spinstructure} and
Section~\ref{sec:canonical}.

\subsection{Gauge bundles}\label{sec:gaugebundles}

We will consider four fields with the fullest correlation structure
envisioned in our conservative constructions, based on the special
unitary groups.  (We will suggest
possible larger constructions with possible enhanced physical interest
in Section~\ref{sec:canonical}.)  The full dynamical structure thus
has structure semigroup isomorphic to $Sp(8,\R )^\C$, and we may think
of our base as the phase space of four pairs of oscillators
complexified, and the function space representative of the base space
being of the form $B \oplus i\circ B$, where $B$ may be, e.g.,
$B = \{ | n_a\rangle \}
\oplus\{ | n_b\rangle \} \oplus\{ | n_c\rangle \} \oplus\{ |
n_d\rangle \} $, a very well behaved spin space representing the
number states of the four oscillators, taken as purely real.  
(We may think of $B$ as initially a Foch
space, but the components will be mixed by dynamical correlations as
the symplectic semigroup runs.  Also, we need not choose energy
eigenstates for a basis.)

Note that the gauge bundles we are discussing in this section involve
active, topologically transitive transformations according to everything we have
done thus far.  Hence, they may change observables (!), unlike the
usual passive gauge transformations (such as one meets in
electromagnetism).  We will ultimately understand them as
transitioning from one ``island of stability'' (e.g., island of
stability within the potentials of the multiple fields which we will
eventually identify as a stable particle) to another.

There is direct mathematical analogy to this structure in the Whitney
sum construction, in which if space $E$ has gauge group $G$, then
$E\oplus E$ has gauge group isomorphic to $G\times G$.  This analogy
follows because the orbits of
$e^{\alpha A}$ and $e^{i\beta B }$ are isomorphic,
$\alpha , \beta \in\R_+$, $A,B \in \mathfrak{g}$, the Lie algebra of
$G$, and we note that
$\mathfrak{sp}(2n,\R )$ and $i\circ\mathfrak{sp}(2n,\R )$ are isomorphic.

Identifying  only ``stable'' gauge transformations,
which will, e.g., take stable states to stable states, one associates
to each ``block diagonal'' subspace not the semigroups
$Sp(8,\R )_\pm$, but the
largest compact sub-semigroup of transformations in $Sp(8,\R )_\pm$,
or $U(4)$.  We arrive at $S[U(4)^\C ]$ from $U(4)^\C$ by any of a
number of routes:
by insisting only on unimodular (unit Jacobean) transformations so that
one avoids (for now)
imputing any physical content to scale changes or inversions of
coordinate orientations, or in order to preserve the normalized probability
measure, or to obtain a simply connected sub-semigroup (which is
thereby really a group), one identifies a representation of the maximal compact
subgroups $S[U(4)\times U(4)]$ as (isomorphic to, according to the
Whitney sum construction) the gauge  transformations for the four
correlated oscillator  system whose
structure (semi-)group was $Sp(8,\R )^\C_\pm$.

The general case of the correct compact
gauge group is deduced from the relationship $U(N)\equiv Sp(2N,\R )
\cap SO(2N)$.  The largest subgroup of unimodular (unit Jacobean),
simply connected group of gauge
transformations is $S[ U(2)\times U(2)]$ for the two oscillator
system.  A similar construction involving three oscillators
will result in an algebra representation with gauge group
$S[U(3) \times U(3)]$, and above we showed that
four oscillators will yield $S[U(4) \times U(4)]$.  The
pattern is obvious.

The full range of these
gauge groups is not available for any given transformation, but they
provide the overall framework for such transformations.  This is because the
transformations are hamiltonian and operate by geodesic transport (see again the
constructions in~\cite{II}, and also the generators of the special
unitary groups act geodesically), and so along any particular
evolution trajectory in the generalized state space
not all quantum numbers can change.  For instance,
there must be some non-zero component along some eigenvector in the
spectral resolution (of the generator of the generalized gauge
transformation) which is non-vanishing under the gauge transformation,
and so
there must be some quantum number which is conserved.  (This is
analogous to saying that $e^{-iHt}\psi$ is not identically zero unless
$\psi$ is orthogonal to {\em all} energy eigenvectors, which requires
that $\psi \equiv 0$ if the energy eigenvectors form a complete
set in our base space.)

Conservatively then, the maximum allowable group of spectrum
generating gauge transformations which are
transitive and dynamical in the case of four fields is
$S[U(4)\times U(3)]$, for three fields, $S[U(3)\times U(2)]$, and for
two fields $S[U(2)\times  U(1)]$, all the result of the constraint that the
 dynamical evolution be geodesic.  See
Section~\ref{sec:breaking} for more details.

In the hamiltonian treatment of electromagnetism, it generally
turns out that the canonical variables do not involve the fields directly, but
the electromagnetic potentials.  (E.g., the transverse component of the
electromagnetic vector potential is gauge invariant and may give rise to
the Aharonov-bohm effect in a region where there are no fields.)  By
analogy, we conjecture that our dynamical gauge field theory describes
the dynamics of the fields in terms of potentials rather than in terms
of the fields themselves.

In electromagnetism, describing the electromagnetic field in terms of
potentials introduces degrees of freedom which are not independent--e.g.,
the electromagnetic fields may remain unchanged by gauge transformations
of the potentials.  The particular choice of a gauge introduces
constraint relations which are used to eleminate the redundant
degrees of freedom.  We thus infer that our theory also has
constraints implicit in it somewhere, but not Dirac type constraints
with arbitrary multipliers.
In particular, our selection of the islands of stability within the
potentials of the field probably figures largely in the constraint
picture.

Some other reminders of the nature of our theory which are distinguished
from classical electromagnetism:
\begin{itemize}
\item{Our theory is phrased in terms of topologically transitive dynamics
rather than a topologically intransitive form, such as the $U(1)$ electromagnetic
gauge theory.}
\item{The canonical variables of phase space do not provide a real Witt
basis, and so the entire spin structure would be dubious if framed in terms
of them.  See also the Appendix in this regard. We require a special type of basis.}
\item{The hamiltonian gauge field theory is framed in terms of occupation
numbers because it is framed in terms of coherent and squeezed states and
creation and destruction operators.}
\item{The lack of manifest covariance is typical of gauge theories--recall
the problems in this regard with $U(1)$ electromagnetic gauge theory.  In the
present context we have established covariant associations via a chain of
isomorphisms--see the following two Sections.}
\item{Recall that there is no configuration space wavefunction for the photon.
It is hoped that by using phase space (spinorial) representation for the
field bosons, those configuration space problems have been avoided.  Recall that
both position and momentum wave functions can be well defined simultaneously
in the rigged Hilbert space formalism~\cite{dirackets}.}
\end{itemize}

The implications of this section should be obvious to anyone familiar
with the
Electroweak and Standard Models.  Whether or nor there is any deep
lesson
here for field theory remains to be seen~\cite{note1}.  It is possible
to construct some
representations of the Poincar{\'e} group in a rigged Hilbert
space~\cite{nagel}.  There is also a construction for relativistic
Gamow vectors~\cite{antoniou,bkwk}.  Given the highly generic
nature of
our methods, we seem to have confirmation that the Standard Model
has the most general gauge structure one would expect from three
fields in the absence of some new and special non-generic
physics, although these gauge groups represent exact symmetries.  
They thus
differ, at least in some details, from the Electroweak Theory and
Standard Model.  We will address the gauge structure further in
Section~\ref{sec:canonical}~\cite{note8}.

\section{Canonical Variables to Spacetime}\label{sec:spacetimevariables}

Starting with a phase space for four pairs of conjugate variables, we
can construct {\em real} probability amplitudes (distribution densities)
over it , as sketched in the preceding Section~\ref{sec:spinbundles}.
We have thus constructed a very well behaved spin representation of
the maximal compact subgroup ($U(4)^\C\equiv U(4)\oplus i\cdot U(4)$)
of the the semigroup of dynamical (=symplectic) transformations on the
complexification of phase space for four pairs of canonical variables
($Sp(8,\R )^\C$), whose action on the
representation space is symplectic (=dynamical) as well.
We have two structures to relate to spacetime.  We have the four
canonical position coordinates to relate to spacetime coordinates, and
we have a non-trivial dynamical structure containing resonances over
the complex extensions (analytic continuation) of our phase space to
relate to a similar evolution structure over spacetime.  The
most interesting subset of this dynamical structure over phase space
is a special unitary group orbit.

The orthogonal Clifford algebra of the
space spanned by the four canonical position
coordinates can be associated
with spacetime structures via the isomorphism $Cl_{O(4,0)} \cong
Cl_{O(1,3)}$~\cite{note3}.
For the dynamical structure, we require a longer chain of
associations.  The largest compact subgroup of $Sp(8,\R )^\C$ is
isomorphic to $U(4)\oplus U(4)$, and we can identify this with a gauge
structure~\cite{note4}.  The Whitney sum rule~\cite{whitney}
and the restriction to unimodular
transformations gives us a gauge group isomorphic to $S[U(4)\times
U(4)]$, which may be identified with $SU(4,4)\cap U(4)$ up to
isomorphism~\cite{knapp}.  We can consider the
restriction of $SU(4,4)$ to $SO(4,4)$, and note we can represent
$SO(4,4)$ in $Mat(2,\BH )$, treating the quaternions as $4\times 4$
real matrices.  But, $Mat(2,{\BH} ) \cong Cl_{O(1,3)}$, and so we see
that $Cl_{U-O} (\widetilde{T^\times \R^4} ) \cap Cl_{O(1,3)} \supset
Cl_{U-S} (\widetilde{T^\times \R^4} ) \cap Cl_{O(1,3)}
\supset S[U(4)\times U(4)]\cap Cl_{O(1,3)} \ne \emptyset $.  We have a
subset of correlated
hamiltonian dynamics over four pairs of canonical variables identified via
isomorphism with a subset of the universal Clifford algebra (geometric
algebra) of a spacetime with local $(+,-,-,-)$
signature.  (More on this point in Section 6 below.)

The $R^{1,3}$ which is the foundation for the $Cl_{O(1,3)}$ above can
be a general riemannian spacetime, and not merely a Minkowski
spacetime~\cite{snygg}, suggesting our formalism is
compatible with general relativity.  (In Section~\ref{sec:canonical}
below we discuss a much stronger link than ``compatibility''.)
By considering the
relatively straightforward chain of textbook isomorphisms above, one
can identify at least a subset of our quantum canonical field theory's
dynamics with ``geometrodynamics'' of a riemannian
spacetime.  What is lacking in the above is any account of hyperbolic
dynamics on either level, or even the demonstration of the existence
of non-trivial connections on this spacetime associated to our
hyperbolic dynamics.  The linkage here needs much more careful
exploration than we will attempt in the present forum.

The nature of the unanswered questions in this formulation is
illustrated by considering the problems posed in describing ``falling
quantum rocks''.  Quantum dynamical time evolution appears to be
geodesic in the
space of states as a consequence of Schr{\"o}dinger's equation.  (But,
recall our caution in attributing group properties to the noncompact
generator orbits, because our Lie algebra may in fact generate
something less than a riemannian, or pseudoriemannian, connection.)
However, the dynamical evolution of a falling rock in general
relativity is along an orthogonal to a geodesic.  Thus, our
isomorphism between the full scope of
quantum dynamics and spacetime geometrodynamics
must certainly map dense sets to dense sets, but apparently need not
necessarily preserve all geodesics, but perhaps only those associated
with stability in the quantum dynamics.  Our chain of textbook
isomorphisms above covered only the islands of stability, and not the
resonances, and is thus  seriously deficient as the source of covariant
dynamics.  It indicates, however, that we should at least be able to
associate our ``particles'' which are coherences of four correlated
fields to extended structures in a non-trivial relativistic
space-time. Naively, we will identify the generators of the compact
transformations identifiable with unitary subgroups of dynamics with
stability and geodesic behavior in spacetime, and the non-compact generators
within the full semigroup of dynamical transformations ``transverse''
to the compact generators will be identified with resonances and
trajectories orthogonal (transverse) to geodesics in spacetime.

There are also conceptual problems
with reductionism in which a subpart of a large system is approximated
as an isolated system, and we should prefer that there be some sort of
analogy, at least, between the process of simplification and
reductionism in our quantum canonical variable dynamics and the
related description in spacetime geometrodynamics.  We have also
shown a possible
linkage between energy-centric Hamiltonian dynamics and mass-centric
general relativity.

\section{``$SU(4)$ canonical gauge gravity''}\label{sec:canonical}

We must acknowledge a bit of ambiguity at the outset.  Conventionally,
one follows Wigner and
thinks of a particle as being described by $\mathscr{P} \otimes
\mathscr{G}$, where $\mathscr{P}$ is a UIR of the inhomogeneous
Lorentz group, and $\mathscr{G}$ is a UIR of
the gauge group.  Issues of fundamental (spin) representations
versus UIR aside, in the preceding
section, we have identified a possible association of our gauge group
to the inhomogeneous Lorentz group directly, suggesting the
possibility that a dynamical representation of $\mathscr{G}$ could
provide a representation of a particle directly in a manner
which is equivalent to the traditional Wigner approach.
Whether such a course will prove physically interesting
is beyond our present scope, and we shall henceforth merely identify our
candidate for $\mathscr{G}$ as ``the gauge group'', adopting a fairly
traditionalist view which respects the traditional definition of a particle.
Given the possible
identification of our present work with a unified version of $U(4)$ gauge
gravity (indicated below), it seems likely there is a fairly strong and broad
identification $\mathscr{P} \Longleftrightarrow \mathscr{G}$, which
requires further investigation.  (Recall that any irrep of the compact
unitary group would be equivalent to a UIR, according to well known theorems.
Thereby, a dynamical representation of $\mathscr{G}$ is equivalent to an
unitary representation of $\mathscr{G}$.)

The intersection of $Sp(8,\R )$ and $O(8)$ is $U(4)=
SU(4)\otimes U(1)/\mathbb{Z}_4$ as a group and $U(4)=
SU(4)\times U(1)/\mathbb{Z}_4$ as a manifold, so that
$U(4)$ has four sheets associated to it,
just as the inhomogeneous Lorentz group does.  
As indicated previously, there maybe means of extending
our well behaved $SU(4)$ structure to larger
structures (in both orthogonal and symplectic topologies), and it is
interesting to speculate that such an extension
may provide a dynamical analogue to PCT, especially given the
associations with spacetime shown in the previous section.  We
identify the Lie transformation groups of primary interest as
belonging to the even sub-algebras of the relevant Clifford algebras,
and thereby identify a spin representation of our unitary group with a
spin representation of some proper subgroup of
the Lorentz group--whether we are speaking of
the connected part only or the full structure depends on whether we are
working with the unitary or special unitary group.

As to the Lorentz
group, we will conjecture that the Equivalence Principle allows us to
interpret the direct sum of inequivalent irreps in our dynamical (spinor)
representation as a sum of particles with correlation.  The natural
interpretation of our spin representations therefore appears to be the
representation that it provides is for pairs of correlated particles,
and is in some sense consistent with Wigner's
definition of a particle as a UIR of the inhomogeneous Lorentz group.
We could extend Wigner's notion of what is a particle to the notion
of a dynamical representation, a
superset of the unitary representations, but it seems natural to respect
his definition in terms of UIR's because thereby in our spinorial constructions
we find we have incorporated the
Equivalence Principle as a by product of the the analytic continuation. This
seems more interesting than regarding a single particle as a correlation
between a pair of field potential structures, since, e.g., a photon in gauge
electromagnetism is a single field potential structure.
We have not made any justification as
to use of the inhomogeneous transformations, however, and so we will
merely be optimistic as to that issue for the present.

As indicated in Section~\ref{sec:focus}, we shall
choose the path of mathematical simplicity for the present and concentrate
on simply connected special unitary groups only, and identify our basic gauge
structure (e.g., possibly modulo dynamical PCT-type transformations) as
$S[U(4) \times U(3)]$, which we may further
simplify to consideration of $SU(4)$ or $ SU(3)$ only.
$SU(4)$ is one of the first quantum symmetries investigated, being
used in nuclear
physics~\cite{wigner1,wigner2,wigner3,hund,franzini,gursey}.
It also figured as a ``spontaneously broken'' symmetry in an
early competitor to the Standard Model (and also a
GUT)~\cite{pati1,pati2,pati3,pati4,pati5,pati6}, has been
explored as a spectrum generating algebra~\cite{bohm6,bohm7,bohm8} (and references
therein), and has relevance to string orbifold
theories~\cite{hanany}.  Our primary concern is {\em dynamically
based} gauge symmetries, the above symmetries are exact, and we abjure
anything ``spontaneous''~\cite{note26}.  (We conjecture that dynamical
transformations which are not elements of the unitary or special
unitary group may break our gauge symmetries, but we really need
greater mathematical justification to make this assertion in
a mathematically proper way .  We are altogether devoted to a
self-consistent dynamics~\cite{I}.)

Our candidate for the full gauge group is of course obtained from the largest compact
subgroup of $Sp(8, \R )^\C$, or $U(4)^\C$, which may be thought of
as $U(4)\times U(4)$ (Whitney sum construction), attracting immediate comparison
to Poincar{\'e} gravity theory (PGT) and the resulting general $U(4)$ theory
of  gauge gravity, obtained from a simple form of the
gravitational Hamiltonian, representing a generalization of the
canonical Arnowitt--Deser--Misner (ADM) hamiltonian form of general relativity.
See, e.g.,~\cite{blagojevic} and references therein, especially chapter 5.  Our basic
dynamical structure reduced to be without correlation would be identified with
the gauge group $U(4)$, so our $U(4) \times U(4)$ theory represents a
correlated, i.e., unified possibly two particle, analogue of the PGT and associated $U(4)$ theory of
gauge gravity.  It is possible to invert the chain of associations and
go from $U(4)$ back to the ADM form of general relativity, and by implication to
similarly go from $U(4)\times U(4)$ back to general relativity as well (with the
added suggestion that we may possibly have explicitly incorporated a
version of the
equivalence principle into our correlation structure, as discussed above.)
Uniqueness in the road back to general relativity is an interesting
and  open question here both physically--in the
context of general covariance--and mathematically.  We desire, of course, that
all physical content be uniquely determined, within the scope of physical
equivalence incorporated in the notions of general covariance, without appeal to
arbitrary assumptions.  Not only is our gauge theory compatible with general
relativity, but there is a fairly well known and well developed parallel in the
$U(4)$ theory of gravity, there is a suggestion of a dynamical basis for PCT,
etc.

There is extensive
discussion of the $SU(4)$ symmetry in the charmed baryons article in
the Particle Data Book~\cite{charm2,charm} and in the quark model section
also~\cite{quark}.  In any event, the basic
16-plet and 20-plet structures, etc., we have pointed to in the Particle Data
Book references should have counterparts in both bosonic and fermionic
representations of $SU(4)$.

All of the above has been largely generic.  If
we were to start with a relativistic phase space, then a similar line of
reasoning to the preceding could possibly
lead to $U(1,3)$ as the principal symmetry~\cite{note0x99}, and
in this case
there is more than one way to look at the dynamical gauge group:
should it be $S[U(1,3) \times U(3)]$ or should it be
$S[U(1,3) \times U(1,2)]$? Perhaps both
are relevant.  In any event, this 
is meant to show quite clearly that there might
possibly be
different gauge groups for 4 generic fields than there will be if the
fields are
specifically identified with spacetime itself.  The
traditional identification of the
gauge theory as a theory of the field potentials rather than of the fields
themselves seems to be in tension with a spacetime based gauge theory in any
event.  Such alternatives to our generic approach could possibly conflict
with the traditional interpretation of gauge theories, while our
generic approach need not.  The lack of manifest covariance in our dynamics
is covered by the multiple ways in which we have illustrated covariant
associations, and also by the already well known isomorphism between
canonical transformations and Lorentz boosts~\cite{kimandnoz}.
The dynamics of four
generic field potentials will originate from the gauge group $S[U(4)\times
  U(4)]$ and the dynamics of spacetime itself might (speculatively) be
viewed as originating (somehow) from
the gauge group $S[U(1,3)\times U(1,3)]$, with the spectrum
constrained as indicated above.  Whether or not we choose to use a Planck
length as a fundamental physical determinant in our dynamics could have
consequences for the
fundamental particles nature is made up of~\cite{note0x100}!  We shall
consider only the generic approach involving a gauge theory of the
field potentials in the sequel, which is a fairly
naive adaptation of the Standard Model to our own notions of
correlated dynamical fields as laid out herein and in the preceding
three articles~\cite{I,II,III}.  It is intended that this article be
thought of as an outline for a more careful program of elaboration and
development.

\section{Bosons}\label{sec:bosons}

We can conjecture what the resulting unified gauge field theory for four generic
correlated fields will be like, by fairly straightforward extrapolation from
the Standard Model, with allowances for the requirements for being mathematically
well defined we have been noting ever since installment two~\cite{II}
where we saw that the incorporation of complex spectra into a quantum dynamics
forced us to depart from the use of unitary transformations for a more
general class of dynamical (=symplectic) transformations, etc.
Our exact $SU(4)$ gauge theory should have a
lot in common with the exact $SU(3)$ symmetry of QCD of the Standard Model, and
we will conjecture that $SU(4)$ is associated with a color chromodynamics of its
own.  In place of the ``three color separation'' of the RGB of QCD, we have a
``four color separation'' we may label CMYK (cyan, magenta, yellow, carmine),
based on analogy the the color separations of the printing industry.  The labels
are, of course, arbitrary.  Being a special unitary group, the even dimensional
irreducible representations are fermionic and the odd dimensional irreducible
representations are bosonic.  We will discuss salient features of the
bosonic representations first:
\begin{itemize}
\item{Associated with the 15 generators of the $SU(4)$ gauge symmetry, we conjecture there
  are 15 color carrying gauge
  bosons--gluons--with zero rest mass and spin 1.  }
\item{Being spin representations of $SU(4)$, our bosonic spinors are a
  direct sum of inequivalent bosonic (i.e., odd dimensional) irreps of
  $SU(4)$.  These irreps need not be UIR's, since they are dynamical,
  a superset of the unitary irreps.  We propose that rather than CUR's we
  should be thinking in terms of CDR's--complex dynamical representations,
  although we will ignore reality/complexity issues for the representations
herein.  Recall that our finite dimensional dynamical representationa are
equivalent to UIR's.}
\item{There are four possible $SU(3)$'s contained in
$SU(4)$, so  it is very likely possible to identify the $SU(3)$ gluons of the
strong  interaction with $SU(4)$ gluons more or less directly, e.g.,
$R\leftrightarrow M$, $G \leftrightarrow Y$ and $B\leftrightarrow  C$, may be
taken as typical of the four alternative $SU(4)$ to $SU(3)$ decompositions.
We envisage two,  three and four color combination particles may be possible,
if we have massles spin one gauge bosons.  The issue of mass is
discussed below, and for now we will discuss a four color QCD.}
\item{Gluonium, glueballs, etc, exist for four colors (and four
anti-colors) just as they exist for the three colors of standard  QCD.  We
expect like colors to repel and color-anticolor to attract,  in analogy to the
electromagnetic charge case. }
\item{The massive $W^\pm$ and $Z^0$ bosons of the
Standard Model raise numerous interesting issues of considerable physical
moment.   We would suggest the $SU(2)$ spontaneous symmetry breaking is
dynamical and not ``spontaneous'', and related to the proposed $SU(4)$ covering
symmetry in some way.  (Massive gauge bosons are discussed below.)}
\item{The massless spin 2 graviton, if it exists,
may be some sort of glueball, or perhaps stem from some special
feature of the  bosonic representation
of $SU(4)$. Note there are possible repulsive color interactions, so there
  may be repulsive field boson interactions attributable to
  the fourth (gravitational) field as well as the attractive
  interaction intermediated by the graviton.   }
\item{In QCD, a significant
percentage of mass of nucleons is  associated with quark-gluon plasma.  Although
the mixed notions of  fermionic quark and bosonic gluon violate our notion of
separation  of perspectives, an identification of a bosonic analogue of the
  quark is indicated below.  We would conjecture that
  all mass is the result of color interactions, principally
  intermediated by
  gluons.}
\item{Correlated bosonic field states in the spin
representations we  have adopted will be identifiable (by isomorphism) with the
form  $\textrm{(boson)} \oplus \textrm{(boson})$, meaning the resulting
representation will be equivalent with an even dimensional  representation, but
this even dimensional representation will not  have fermionic exchange
properties since it is entirely associated  with a different bilinear form (and
associated scalar product).  We  might call this a pseudo-fermionic
representation of the basic gauge  structure, i.e., correlated bosons may appear
to be single fermions  if exchange properties are not carefully dealt with.}
\item{In our dynamical treatment, we will find only bosons.  There is
  no proper place for quarks in this dynamical perspective, although there may
  be bosonic structures related to them (see below).  The bosonic and
  fermionic $SU(4)$ representations need to be studied for their
  structure, and there should be similarities and correspondences
  between many, if not all, of the structures in the respective
  bosonic and fermionic perspectives.}
\end{itemize}
As indicated in the last item above, the need for rigorous separation
of perspectives means that our chromodynamics will differ
significantly from the QCD associated with the Standard Model, and not
just in the addition of one more color or
with the attendant possibility of four color combinations in addition
to two and three color combinations already familiar in QCD.  For
instance, the notion of a ``quark-gluon plasma'' seems an oxymoron,
given the enforced separation of bosonic and fermionic perspectives
(based, in part, on dimensionality of the irreducible representations
our spinors are built up of).  This seems to indicate the broad outline
of our conjectured extrapolation from the Standard Model as to our generic
hamiltonian four field gauge theory.

\section{Symmetry Breaking and Bosonic Mass}\label{sec:breaking}

Mass production in the Standard Model is due to what is there called
spontaneous symmetry breaking and which we have preferred to call
dynamical symmetry reduction.  There are several implications to our
construction which deserve at least speculative comment.  Recall that
the $U(N)\times U(N)$ structure arises (via isomorphism) from
dynamical correlation being introduced in a specific way, and the
spectrum comes from the Lie algebra of the generators of
$S[U(N)\times U(N-1)]$.  One kind of correlation
is binding, and binding is associated with negative energy, and after
analytic continuation our energy spectrum is unbounded from below, at
least formally.  One {\em possible} implication of this construction
is that the Big Bang may have been energetically neutral, with
positive mass and energy balancing the negative correlation energy.
The $U(N-1)$ in the $S[U(N)\times U(N-1)]$ might therefore be regarded
as the postive mass part, being associated with a $U(N)$ of massless
bosons in correlation, e.g., the $W^+$ particle is paired with various
massless field intermediary particles (photons, gluons, gravitons,
etc.) such that the total energy of each part is equal to that of the
other part and the correlation effectively means they are equal and
opposite to each other.  This is of course only a speculative explanation
of a possible basis for mass production, the equivalence of mass and
energy, etc.  

In the remainder of this subsection, we will explore
other other possible implications in a pretty naive manner in the
interest of brevity--we are attempting a fairly straightforward
extrapolation from the Standard Model's results to our
constructions, and we make many speculations along the way to 
indicate issues in need of resolution. The point emphasized here 
is that the correlation need not be between matter and other matter,
or between matter and antimatter, for instance, but might even serve as
a source of particles in and of itself: the correlated potentials
have the symmetry of the gauge group, and the spectrum arises
from this symmetry which thereby provides the source
of particles and antiparticles. One example of correlation of the type 
we refer to is bonding involving an attractive potential, which can result
in the release of other forms of energy due to overall energy conservation.
(Typically this is a kinetic energy release interpreted as a temperature
rise, such as in chemical bond formation.) We will evade any efforts
attempting to force us to specify particle formation mechanisms at
this point. There are deep issues here best left for later.

The first hurdle to be faced in identifying the spectrum is to
identify the fundamental symmetry of interest.  As indicated
previously, the basis symmetry for these islands  of stability within
our hamiltonian dynamical field-potential constructions is $U(4)\times
U(4)$,
but  we are working with spin manifolds, and spin manifolds have
simply  connected fundamental groups--indicating that in this case
mathematical conservatism should restrict us to special unitary
groups.   Since we make no present attempt to extend the notion of a
proper spin manifold, we shall consider $S[U(4)\times U(4)]$ as our
simply connected fundamental group, and serves as our gauge group underlying
these  constructions.  Now $U(4)\equiv U(1)\otimes
SU(4)/\mathbb{Z}_4$,  which we can think of in several ways, including
$U(4)\longrightarrow \left\{ U(1)/\mathbb{Z}_4 \right\} \times SU(4)$
or  $U(4) \longrightarrow U(1) \times \left\{  SU(4)/\mathbb{Z}_4
\right\} $, substituting a direct product for the tensor product.
Proceeding naively, we can think of $S[U(4)\times U(4)]$ as
$$
S\left[ \{ (U(1)/\mathbb{Z}_4)_{PCT} \times SU(4)_{CMYK} \}
     \times \{ U(1) \times (SU(4)/\mathbb{Z}_4 ) \} \right]  .
\nonumber
$$
Because of the previously demonstrated injective embedding map
$\varphi : U(4) \hookrightarrow \mathscr{P}$, in the above
decomposition  we have conjecturally identified the
$(U(1)/\mathbb{Z}_4)_{PCT}$ term with dynamically based PCT
transformations, and will henceforth ignore this factor.  Because of
this  mapping (demonstrated in Section~\ref{sec:spacetimevariables}),
$SU(4)_{CMYK}$ 4-colored particles have covariant associations which
we  will conjecture is what makes them real and observable.  This
reality/observability should extend to subsymmetries of our
color-$SU(4)$, but not to the so called ``fundamental particles''
argued for in Section~\ref{sec:hyperbolic} below--these, we would
argue, are  real in some sense but lacking in covariant associations
which make them a ``particle'' at this level of development~\cite{note0x100}.

Beginning with the spinorially well defined part of $U(4)\times U(4)$,
our fundemental symmetry has how been simplified to
\begin{equation}
S[U(4)\times U(4)]\Longrightarrow S\left[ SU(4)_{CMYK}
     \times \ U(1) \times (SU(4)/\mathbb{Z}_4 )  \right] \quad .
\label{eq:reducedgauge}
\end{equation}
The spectrum actually observed comes not from this, but, as indicated
earlier, from the reduction of this by one generator to allow for the
conservation effects of parallel transport, i.e., taking out the
$U(1)$ factor, for instance.  Our basic gauge group is now
$SU(4)_{CMYK} \times (SU(4)/\mathbb{Z}_4 )$.  There are 4 copies of
$SU(3)$ in $SU(4)$, so we can naively identify the
effective dynamical gauge group as
$$
SU(4)_{CMYK} \times SU(3)_{Flavor}
\label{eq:finalapproxgauge}
$$
One can also speculate about a symmetry reduction
$SU(4)/\mathbb{Z}_4 \longrightarrow SU(3)/\mathbb{Z}_4$ to conjecture the
possibility of there being 4 generations among the 8 generators of
$SU(3)$, an issue primarily of interest to fermions, see
Table~\ref{table:1},  e.g., thinking in terms of 4 pairs, or a total
of 8, quarks being associated with the $SU(3)$ symmetry.

It is recognized geometrically the process of ``spontaneous symmetry
breaking'' is in fact reduction of a principal bundle.  Because of our
preoccupation with dynamical structures, and because our principal
bundle has as its group the symmetry transformation relating to
compact  dynamical transformations, we much prefer the term
``dynamical  symmetry reduction'' as being both physically more
precise and also mathematically more precise.

We thus expect 15 (probably massless)
gauge bosons associated with $SU(4)$ and 8 possibly
massive gauge bosons associated with $SU(3)$.  Note that there may not
be any analog to the Goldstone theorem restricting the masses of the
$SU(4)$ gauge bosons, and
that if there are any analogues of the Higg's boson they would be associated
with the $SU(3)$ symmetry and could be vastly more massive than the Higg's
associated with the $SU(2)$ symmetry.  Renormalizability is
mathematically sufficient for spontaneous symmetry breaking, and our
very well behaved (see installment one~\cite{I}) wavefunctions should
be renormalizable, if not indeed already ``renormalized'', so
mathematical processes equivalent to spontaneous symmetry breaking
should follow in our formalism, with the attendant massive gauge
bosons resulting.

\subsection{Spectrum generating algebras}\label{sec:spectrumgenalg}

It would appear to the author--who could probably be said to only
understand sufficient of the matter to be dangerous--that from the
perspective of spectrum generating algebras it is possible to recover
the Standard Model from our constructions, at least in large measure.
We note further that a representation of electromagnetic current and
electromagnetic charge generators is known for $SU(4)$~\cite{bohm7}.
Perhaps a representation for the graviton can be found as well.  Also
of interest that there are different fermionic and bosonic discrete
subgroups to $SU(4)$~\cite{hanany}, suggesting that there may be
observable choices to the choice of topology in addition to those
alluded to back in Section~\ref{sec:tentative}.  We would conjecture
that if one consistently follows a perspective, then any possible
observables of that perspective should be observed in context.  The
context relevant to a perspective might be shaped by such considerations
as energy scales (e.g., Compton vs. de Broglie wavelength), etc.

The spectrum generating algebra approach evolved from the ``spectrum
generating group'' approach, also referred to as the dynamical group
and as the non-invariance group.  The present series of papers has
adopted the approach well known from classical nonlinear dynamics that
the dynamical group is the relevant group of symplectic
transformations (also known as the group of area preserving maps in
the case of simple systems).  We have thus extended this notion from
classical hamiltonian mechanics to the arena of hamiltonian quantum
fields.

\section{Fermions}\label{sec:fermions}

In the fermionic perspective, we adopt quark flavor as the smallest fermionic
analogue to the bosonic color charge, e.g., the fermionic counterpart of the two
color gluon, the two quark meson, is constructed from the four udsc quark
flavors in direct analogy to the role of the four bosonic cmyk colors in the
gluon construction, suggesting the possibility that an identification between
structures in the alternative perspectives exists.  On general principles, it
seems like there should be some sort of identification between bosonic color and
fermionic flavor, but this tantalizing identification really does not
inescapably mean (at this level of development) that gluon=quark or
glueball=meson.

The fermionic effects of the dynamical apparent symmetry reduction can be predicted for
$SU(3)$ and its eight generators, which we associate with quarks in the usual way.  This is best visualized by the following Table~\ref{table:1} which extends a familiar table to a fourth column, reflecting the addition of a fourth generation.
\begin{table}
\begin{tabular}{|c|c|c||c|}
\hline
\multicolumn{4}{|c|}{Quarks} \\ \hline \hline

I & II & III & IV \\    \hline
u & c & t & i \\    \hline
d & s & b & o \\    \hline \hline

\multicolumn{4}{|c|}{Leptons} \\ \hline \hline

$\nu_e$ & $\nu_\mu$ & $\nu_\tau$ & $\nu_\zeta$ \\ \hline
e & $\mu$ & $\tau$ & $\zeta$ \\ \hline

\end{tabular}

\caption{Defining fermionic particles from the gauge group generators for $SU(3)/\mathbb{Z}_4$.  It is assumed that a process equivalent to spontaneous symmetry breaking has led to mass generation.   There is an additional pair of quarks (``inner'' and ``outer''), a fourth neutrino, and a fourth lepton (which we call the ``zeta''), all associated with with the massive $SU(3)$ symmetry.  The $b'$ fourth generation search information in the Particle Data Book are relevant to the $i$ and $o$ quarks conjectured above.  The $\zeta$ and $\nu_\zeta$ can only be expected to arise directly in significant numbers during the course of very high energy phenomena in nature, e.g., supernovae and the like, so the lack of observation of any fourth neutrino such as the $\nu_\zeta$ is not yet a criticism of the above predictions, but also indicates these may be very hard to confirm experimentally.  If the mass of the $\nu_\zeta$ is assumed comparable to that of the three known neutrinos, there may be some effect on neutrino oscillations, which is probably the easiest place to look.}

\label{table:1}
\end{table}

Recall that underlying the choice of alternative bosonic or
fermionic representations lies merely the choice of topology in which to
complete Cauchy sequences.  It is doubtful whether we can experimentally
determine the actual topology of our spaces of states, and mathematically we
do not expect the choice of topology to affect scalar products, so we very
strongly expect that there should be some strong sense of equivalence, or at
least an identification, between the observables of the two alternative
perspectives.  (In Section~\ref{sec:tentative}, we indicated how the
choice of topology may be relevant to mass and energy scales, etc.,
and the issue of wave--particle duality.)  It is the matrix elements
that matter, and the prespective chosen should not matter as to
these--topology should not change the integrals, etc.~\cite{narici}.
At a minimum, a construction as outlined in the preceding section does
seem to set the basic observables in each perspective, and establish a
sort of invariance as to those
basic observables using  the mass-energy equivalence well known from relativity.
If we take this model literally, we would expect to find some expression in
nature of both the bosonic and the fermionic perspectives, meaning
when  we adopt a particular perspective as relevant to a particular
experiment, we expect to find events in nature which may be
interpreted consistently in that perspective.  The unresolved mystery
at this point of development is whether or not there is any strong
identification, perhaps even equivalence,
between any of the structures we may find in both perspectives.  More
issues to resolve in the future.

This suggests an
identification between energy centric gauge bosons associated with
hamiltonian dynamics, and mass centric gauge fermions associated with
orthogonal (Riemannian or pseudo-Riemannian) geometry.  Both types of
spinors are associated in our construction with representations of
unitary groups only, and
any  identification could not be extended to all
of dynamics, for instance.  This identification is between fermionic
and bosonic stable structures (topological invariants?) , and does not
extend to resonances.  Whether we regard an elementary particle as a
stable bit of geometry or as a stable coherence of fields
depends on our choice of topology for the representation of that
particle (and
the consequent even or odd dimension of the representation).  We
have been taught to expect that the choice of topology in our space of
states should have no observable
consequences, absent a Cauchy (infinite) sequence of physical measurements.
We would assert this principle applies to the mathematical
representation of the system rather than acting as a
constraint on the system itself: one must choose an
appropriate vocabulary.  If one adopts a vocabulary to make a prediction,
then one's results must be analyzed in a consistent vocabulary, but
the model is not the thing itself.

The ``islands of stability'' in either perspective, i.e., the ``elementary
particles'',  should in some sense
be invariants, characteristic structures in our space(s)
of states.  A general gauge field theory is associated with the
potentials of the gauge field, just as in the U(1) electromagnetism
gauge theory, and for us the islands of stability
represent topological features of the multiple field
potentials which are associated with stability
and it should not matter whether we think of them as a glueball or
as a concatenation of quarks.  The fermionic or bosonic perspectives
represent a choice of toplogy and that choice should not be relevant
for features resolvable in both topologies--however, there are
different numbers of discrete fermionic and bosonic representations of
$U(4)$~\cite{hanany}, for instance. This means correspondences
between ``particulate lumps of matter'' and ``particulate intermediaries
of dynamics'' will not be $1:1$ correspondences in general.

Resonances are another matter!!!
Although we should
probably regard resonances as ``particles'', we have shown in installment
two~\cite{II} their intimate association with dynamics making
representation of
them by fermionic spinors physically and mathematically
problematic--they are the essence of non-trivial dynamics. The
preceding chapters have
shown it is possible to associate bosonic and fermionic
representations of the unitary
group with fermionic representations of the Lorentz group, but whether or not
there is any such identification extending to $Sp(8,\R ) \backslash U(4)$ is
unknown.

There are a lot of intriguing hints of how things should work out when an
exact four color QCD is fully developed with strict separation of
perspectives.  Elaboration of these are, however, major undertakings
which we will postpone completion of for another date and another
forum.  We have gone on long enough, so will conclude with a
couple of sections making some related additional predictions and
also in formally delimiting our work in various ways.  We have
identified our dynamical structures with a quantum field theory
identifiable with certain features of particle physics because the
particle formalism seems (in the author's view) to offer the most
immediate prospects for physical confirmation.  The mathematical
structures, and to some extent their physical interpretation, seem to
be largely generic, however, and we thus anticipate that the preceding
separation of unitary and hyperbolic dynamics should extend to any
fluid with long range correlations, plasmas for instance .

\section{Whence Hyperbolic Quantum Dynamics?}\label{sec:hyperbolic}

According to Wigner, particles are UIRs of the inhomogeneous Lorentz group.
Whether or not we should think of resonances as particles depends on
whether we identify a particle as $\mathscr{P} \otimes\mathscr{G}$
(where  $\mathscr G$ is the gauge group) or allow
dynamical semigroups $\mathscr{D}_\pm$ for which there exists a map $\phi$ of
the type indicated in the preceding section such that $\phi\cdot\mathscr{D}_\pm
\cap \mathscr{P}\ne \emptyset$, as well as the particular choice of dynamical
semigroup $\mathscr{D}_\pm$ itself.  There are very well behaved
semigroup representations of the inhomogeneous Lorentz group~\cite{nagel},
and so to identify a resonance as a particle, what is needed is
to associate the dynamical semigroup with a representation of the
inhomogeneous Lorentz semigroup (in the sense of an extension of the above
$\phi\cdot\mathscr{D}_\pm \cap \mathscr{P}\ne \emptyset$).  As a third
alternative, one might extend Wigner's definition of
a particle from the elementary particles $\mathscr{P} \otimes \mathscr{G}$
to $\mathscr{P}\otimes \mathscr{D}_\pm $, where $\mathscr{D}_\pm$ is the
dynamical semigroup containing the compact gauge group $\mathscr{G}$
as the maximal  compact subgroup.
This last course seems perfectly reasonable also, and in the present
context  concerning four fields we can identify $\mathscr{D}_\pm = Sp
(8,\R )^\C_\pm$  and
$\mathscr{G} = S[U(4) \times U(4)]$, thereby expressly
extending the notion of a particle to include resonances,
but the first alternative is also interesting and will be explored below.

There is some degree of concordance between spacetime geometry and the
non-trivial
dynamics of quantum fields as previously indicated when we showed the
links between the unitary Clifford algebras and the Clifford algebra
of spacetime.   That association concerns equilibrium and invertible
dynamics.
There are also  links relative to that sub-aspect of all possible
dynamics which is identifiable with semigroups and hyperbolic dynamical
evolution, implying that hyperbolic dynamical evolution (including
non-time-reversible dynamics) is possible in spacetime just as it is
in the hyperbolic  hamiltonian dynamics  of our canonical variables.
A complete  quantum dynamics must be able to account for the falling
quantum rocks previously alluded to - the resonances which are transverse to
the stable states.  Note also that there may be
some possible aspects of even non-trivial dynamics which are not associated
with influencing the geometry of spacetime (e.g., wholly internal
irreversible transitions of a resonance, at least in toy models and
gedanken experiments), and we must allow for possible aspects of
spacetime geometry which do not have
any non-trivial dynamical significance (e.g., stable or strictly periodic
phenomena). This requires, for instance, that our cosmological thinking ought to
include consideration of hyperbolic dynamics such as arise in non-conservative
and/or open cosmologies, should we consider the wavefunction of the
universe.  Possible conservative, closed or cyclic
cosmologies seem excluded by the experimental observations of accelerating
expansion~\cite{perlmutter,garnavich}, and seems also in tension
with the observation of (near) flatness on cosmological scales of distance.

The basic mass spectrum for observed particles seems to be deducible
from the preceding section's development of a generic hamiltonian four
field gauge field theory, and we shall adopt an optimistic attitude
concerning the outcome of those as yet unmade calculations and as yet
unmade experimental tests.  This
seems reasonable because there seem to be relatively clear principles
for extrapolating from the Standard Model to a hamiltonian quantum
field theory along the lines we advocate for any number of fields, and
in particular to four fields, thereby adding gravity to a modified
Standard Model.  (The principal differences from the Standard Model
are conceptual and
seem to stem from the separation of bosonic and fermionic
perspectives, based on mathematical necessity.)  There are further
indicated tests for determining whether the additional fourth field
should be thought of as ``generic'' or if one should think in terms
of a quantized spacetime.

Our driving principle throughout has been to obtain a probabilistic
description of hamiltonian correlated dynamics which is as well
defined as the author knows how to make it.  We have a core dynamical
construction which is well defined over compact dynamical
transformations--we can represent compact dynamical transformations in
a well defined spin geometry, associated to well defined spin bundle
structures, etc.  Representing the generators of the compact
infinitesimal  dynamical transformations are bosonic spinors, and
associated to them are representations of the same group structure by
fermionic spinors.  Those fermionic spinors are associated with
particle representations (either in the sense  of $\mathscr{P} \otimes
\mathscr{G}$ or $\theta\cdot \mathscr{G}\cap \mathscr{P}\ne
\emptyset$)--they are creatures of Riemannian geometry (symmetric
metric), so that we can think of their velocity and position as
sufficiently localized and well defined (probabilistically) in a
way that says ``particle''.  There is no comparable sense of ``this
is a particle'' in quite the same sense in the symplectic geometry
of hamiltonian dynamics (the uncertainty principle notwithstanding).
Conversely, there is no way you can speak with mathematical precision
of the hyperbolic dynamics of fermions.  Recall how in installment
two~\cite{II} we used conjugation by a semigroup element of a
generator to obtain a Breit-Wigner resonance--that conjugation was
generated by a hyperbolic generator.
It is natural to think of a stable particle as a lift of another
stable particle wave function--in our finite dimensional
representation of our compact gauge group, we can associate a
family of particles with a single representation in this way.
In the sense of installment two~\cite{II} then, interparticle
interactions appear as multiplication (or conjugation) by a group
element, using the sense of symplectic action and duality indicated
in installment two.  In the case in installment two ~\cite{II}
where we were looking at the expectation of the Hamiltonian, $H$,
we saw how the conjugation changed the Hamiltonian, i.e., changed
the interaction, etc., in addition to the changed wavefunctions.
The question emerges, why should we not consider the
infinitesimal generators of $\mathfrak{sp}(2n,\R )^\C_\pm
\backslash \mathfrak{u}(n)$ as providing labels for particles
as well?  Although this may possibly require acceptance of
resonances as particles, we know that resonances exist and that
whatever produced them cannot have anything to do with
$\mathfrak{u}(n)$ and gauge bosons, e.g., there are actually
existing in nature ``fundamental'' particles which are not
elementary particles in any accepted sense of what an elementary
particle is.  These ``fundamental'' particles are the resonance
makers--they transform stable particles into resonances.

You cannot extrapolate compact dynamics into non-compact
dynamics using only compact dynamics as a tool.  The
compact dynamical transformations form a closed and
compact group--there can be no noncompact dynamics, and
thereby no resonances, in any $U(n)$ elementary particle
theory or which may be obtained from such a theory in any
finite limit.  There is not any finite limit (macroscopic
or otherwise) whereby you can go from gauge bosons to
hyperbolic dynamics of any kind.  Complexity won't do
the job--complex dynamics obtained from compact dynamics
is merely complicated compact dynamics, and can not rise
to parabolic or hyperbolic dynamics.  Repulsive color
interactions which are a part of compact dynamics cannot
yield non-compact dynamics in any finite limit, such as in
any finite universe.  You might adopt a holistic approach
to nature, but finite causal horizons say there can only be
a finite number of interacting particles--if there are finite
causal horizons like the Big Bang, you cannot get to
hyperbolic (or parabolic) dynamics by the necessary 
transfinite means through
invoking them.  Even speed of light exceeding accelerating
expansion, such as may have occured in the early universe,
if it is of finite speed and duration will still produce finite
causal horizons.  Tachyonic matter, if it exists, is causally
unrelated to ordinary matter.  There seems to be no
mechanism for producing hyperbolic dynamics, and yet
resonances exist both microscopically and in form of the
expanding universe as a whole--so that gauge field theories
must provide an incomplete description of the dynamics of
elementary particles, and therefore an incomplete
description of nature.  It would seem that the group
description itself is too restrictive, and you must include
dynamical semigroups in order to have  a sufficiently
rich description of nature to accomodate the existence
of resonances.  We shall conjecture the existence of
particles associated with the dynamical semigroup but
outside of  the maximal compact subgroup, initially
calling them ``fundamental particles'', noting clearly
they are not ``elementary particles'' in any accepted
sense of that familiar term, and also because their description is constrained
to symplectic geometry, meaning they are representable
as bosons only, but not as gauge bosons.  (We will adopt
the name chimeric for these bosons in the next section,
for the reasons given there.)

This observation that there must be fundamental particles
 which are not elementary particles (gauge bosons)
has a number of important possible consequences.
Most noticeable is the larger number of particles
which emerges overall, as summarized in Table~\ref{table:2}.
\begin{table}
\begin{tabular}{|c|c|c|c|}
\hline
Fields & Generators & Elementary Particles & Fundamental Particles \\  \hline
2 & 10 & 3 & 7 \\    \hline
3 & 21 & 8 & 13 \\    \hline
4 & 36 & 15 & 21 \\    \hline

\end{tabular}

\caption{Defining particles from the dynamical semigroup
generators.  The ``fundamental particles'' are conjectured
bosons associated with noncompact dynamics, and their
interaction with an elementary particle would be the
formation of a resonance.  They may be possible sources
of both Dark Matter and Dark Energy--they may be massive
just as gauge bosons may be massive.  Their conjectured
existence makes large mathematical assumptions if they
are to be accepted as well defined mathematially. We will call them chimeric 
bosons.}

\label{table:2}
\end{table}

Thus, in the Standard Model $SU(3)\times SU(2) \times U(1)$
adopted to our view, there could be seven additional
(probably massive) fundamental bosons in addition to
the $W^\pm$ and $Z^0$ associated with the $SU(2)$ gauge
symmetry, and another thirteen additional fundamental
particles associated with the $SU(3)$ symmetry.  (These thirteen
would probably not be massive in this context.)  Our four field
generic gauge gravity would be based (modulo PCT-type transitions)
on $SU(4) \times SU(3) \times U(1)$, with eight massive gauge
bosons (possibly regarded as including the $W^\pm$ and $Z^0$)
and another thirteen possibly massive fundamental bosons,
all associated with the $SU(3)$ symmetry, plus fifteen elementary
and an additional twentyone fundamental (and presumably massless)
bosons associated with the $SU(4)$ symmetry.  These emerge from
our constructions when the notion of gauge symmetry reduction is
extended to the full dynamical semigroup, and provide obvious
candidates for Dark Matter and Dark Energy.  It is hoped that
the reader will at least receive this as a principled speculation
worthy of further exploration.  Such fundamental particles are
mathematically well defined only if we can extend our spinor
structures to the full dynamical semigroup, and can be massive
(according to analogy to present understandings of mass mechanisms
for gauge particles) only if we can speak coherently of something
analogous to principal bundle reduction for ``semigroup bundles''.

\section{Covariant Identification of Particles}\label{sec:covariance}

The relationship between the symplectic transformations and
Lorentz transformations for photons in explored in chapter 7
of~\cite{kimandnoz}.  A Lorentz boost is equivalent to a symplectic
transformation, and vice versa, in the case of photons, and a
similar equivalence should hold for elementary particles associated
with $U(4)$ type gauge structures.  As indicated before, we
have adopted the position that all particles are minimum uncertainty
eigenstates of various fields, and that the transformations
between eigenstates correspond to squeezing transformations in
the language of quantum optics, and to transitive canonical
transformations in the language of hamiltonian dynamics.  In the
context of chimeric bosons, 
a general symplectic transformation need not be identifiable with 
a Lorentz boost, meaning that in general the difference between
one particle and another may not be covariant: different observers
may assign different quantum number identifications to the same
particle, different participants to the same process, etc.,
according to our constructions (and this seems to be a
generic problem of hamiltonian mechanics!).

If the quantum numbers which characterize a particle are derivative
of the $U(4)$ gauge symmetry, then they have covariant associations
as indicated preceding.  Thus, all observers should be able to
agree that a given particle is an electron or a $W^+$, or whatever.
The same cannot be said of the quantum numbers which we associate
with the chimeric particles described in
Section~\ref{sec:hyperbolic} which are ``fundemental'' but not
gauge--these particles are chimeras in the sense that different
observers might identify different quantum numbers for them,
and so our preferred name for them is chimeric, derived from
analogy to the chimeras of ancient mythology.

We therefore suggest the name chimeric bosons because of
this lack of covariant association. Like all of the states we
have dealt with in this series of papers, they are minimum
uncertainty eigenstates, but I am not prepared to discuss them
much more than by simple analogy to the squeezed and coherent states
of quantum optics, also obtained by the application of symplectic
transforms to Foch states.  See, e.g., \cite{kimandnoz}, especially
chapter seven, where the relationship between symplectic transforms
and Lorentz transformations is explored in the case of photons--a
Lorentz transformation is equivalent to a symplectic transformation
(squeezing transformation in the language of quantum optics) in the
case of a photon.  In the present context, that symplectic
transformation might work an apparent transformation of a chimeric
boson into another type of particle, since in the case of chimeric
particles there are no covariant associations permitting universal
identification of its fundamental quantum numbers. Those
non-equilibrium squeezed photon states have sub-Poisson statistics, as
we have shown our own resonant states have.  Our analogy would be
to suggest that one observer might say p-squeezed and another say
q-squeezed, to use the jargon of squeezed states in quantum optics.  
From our development of these notions from foundational
notions of dynamics
and the observation that the islands of stability in four-field
dynamics have covariant associations, and the observation that stable
bosons are associated with the generators of transformations between
the islands of dynamical stability, we are led to conclude that there
must be bosons associated with the carrying of our four fields of
force which are responsible for the accelerating expansion of the
universe (among other effects) which lack unique covariant
identification.  However highly we esteem it, general covariance
is at odds with the hyperbolic dynamics implied in 
the accelerating expansion of the universe, or there
are more than four forces in the universe, or we have some fundamental
misunderstanding of hamiltonian dynamics, etc.

The author hopes
that at the very least he has demonstrated the presence of a
foundational crossroads and has endeavored to offer a resolution.

\section{Possible Quaternion Structure?}\label{sec:possible}

In this section, we tie up some odds and ends.

We began with a pair of spaces in direct sum, each element of which is
a sum of states associated with four correlated oscillators.  If we
were to allow full expression of the
correlations possible between four oscillators,
each of the states for four
correlated oscillators would have a quaternionic structure.  We have
also made use of an isomorphism involving $Mat(2,{\BH} ) \cong
Cl_{O(1,3)}$, further suggesting a possible
quaternionic structure.  We will
not make further investigation of quaternionic structures, and
dynamics with quaternionic correlation structures, in this forum, but
will note in passing that quaternionic quantum mechanics is a fairly
mature field, with many points of interest. See, e.g.,~\cite{adler}.
Our reason for avoiding quaternions
is that quaternionic notions would raise tensions, if not
outright conflict, within our fundamental structures.

The equivalence principle of Einstein is related to Mach's principle,
and all that is required for the equivalence principle (``your can't have
mass $A$ without also having a mass $B$'', to paraphrase Wheeler's
well known version) is that $| A\rangle \oplus |B\rangle$ lie in a
space wherein there is a correlation between the $|A\rangle$ and
$|B\rangle$ components.  A Foch space structure alone will not
suffice for the desired correlation.  In the present context, we
envision a complex symplectic
structure between $|A\rangle$ and$|B\rangle$ components, reflecting
the real form of our correlated dynamical semigroups,
$Sp(2n,\R )^\C_\pm$.  (A real symplectic structure might possibly
suffice, however,
or, indeed, even an orthogonal correlation).  In light of the
quaternion issues with respect to $|A\rangle$ and $|B\rangle$
suggested above, if $|A\rangle$ and $|B\rangle$ were quaternionic then
the existence of a complex symplectic structure
(complex correlation) between them would raise the possible specter of
non-associative octonionic structures, with consequent loss of
functoriality of mappings, categoriality within the meaning of the
mathematical theory of categories, and ultimately even the boolean
structure underlying our probability
interpretation~\cite{cox} vanishes.  We emphatically do not dismiss the
quaternion and biquaternion approaches outright, but indicate there are
possible problems if they are not used carefully and leave such issues
for another day.

\section{Concluding Remarks}\label{sec:remarks}

The formalism we outline in this series of articles
makes no assumptions about the nature of
particles, their numbers or their conservation, only that there were
some structures on phase space which could be localized in some sense
(so that our measures are finite)
and which possessed non-trivial dynamics~\cite{note6}.
Since we are working with distributions, dynamical evolution is by semigroups,
and there is an intrinsic dynamic arrow of time to the formalism (subsequent to
the analytic continuation, which physically introduces correlation
into our
considerations, and thereafter mathematical necessarity places us in the
distributions).  Conversely, this arrow of time requires that our quantum
desriptions make use of distributions and the rigged Hilbert space
structure~\cite{qat1,qat2,qat3}, such as we use here.  Since we see
no evidence for stationary or strictly
cyclic cosmological dynamical behavior (in fact, quite
the contrary), and since we do see expansion, we must expect that the
rate of expansion increases exponentially (hyperbolically)--as
apparently it actually does~\cite{perlmutter,garnavich}.
We have also illustrated
how these hyperbolic dynamical system
results may be obtained from a quantum description of the
universe.

Because our results have touched on numerous foundational issues,
including the universal validity of general covariance (!!), it is
perhaps appropriate at this point to engage in sort of an executive
summary of the steps making up the developments we have undertaken,
and differing substantially in its outlook from the summary in
Section~\ref{sec:nontrivial}.
While the author has not been as fastidious with his mathematics
throughout as a mathematician would have been, he has tried to
 exercise much greater mathematical care than is typical in physics,
and we have repeatedly seen physically interesting notions emerge from
greater mathematical care--this is lesson number one.

In the first installment~\cite{I}, we made a recapitulation of the
rigged Hilbert space formalism and
\begin{itemize}
\item{Represented classical hamiltonian dynamics on phase space with
probability amplitudes rather than point localizations.  Because
both position and momentum are simultaneously defined variables
in the rigged Hilbert space formalism, there is no need to resort
to Wigner functions or quasi-probability distributions, or the like,
and one may work represent quantum dynamics on phase space directly.}
\item{Showed how the analytic continuation of this hamiltonian
probabilistic dynamical description is indistinguishable from quantum
theory in the rigged Hilbert space generalization of the
Schr{\"o}dinger theory, with distributions and an arrow of time.  Our
probability amplitudes are generalized gaussian wave packets, of
mathematical necessity.}
\item{Showed how physically, the analytic continuation represents the
  introduction of a type of mathematical correlation.  Recall that
  with analytic continuation, the energy spectrum may be unbounded
  from below--e.g., the correlation introduced may in some sense be
  identified with binding.}
\item{Explored the multiple types of scalar product which can be
  defined on phase space, anticipating the bosonic and fermionic types
  of scalar products which emerged in our later constructions.}
\item{Showed that the mathematics was consistent with a field theory
  interpretation, i.e., the formalism really does not distinguish
  whether a particle is a ``marble'' or a localized excitation of fields.}
\end{itemize}

The second installment~\cite{II} demonstrated that the incorporation
of resonances with their complex energy eigenvalues required us to
meet a lot of mathematical and physical needs by using multicomponent
vectors (spinors) and a novel definition of the adjoint:
\begin{itemize}
\item{We required esa operators in order to represent the canonical
  Lie algebra inclusion $\mathfrak{g}\subset\mathfrak{g}^\times$, so
  that, e.g., the generators of the two algebras are identified with
  the same tangent vector to a geodesic.  A consequence is that the
  spectral theorem applies, and we have well defined spectra.  This
  suggested that the adjoint should be an algebra involution for the
  Lie algebra of dynamical observables--e.g., for the Lie algebra of
  the group of dynamical transformations, and the largest possible
  and most general dynamical group is generally accepted to be
  the group of symplectic transformations.}
\item{In order to incorporate complex spectra, we were required to
  adopt the time reversed scalar product if the Hamiltonian was to be
  the generator of dynamical transformations.}
\item{For the adjoint operation to be a well defined algebra
  involution, we were required to adopt for our ring of scalars the
  commutative and associative real algebra (with two units) $\C (1,i)$
  rather than the field $\C$ as has been the custom heretofore.}
\item{We were careful to respect the Lie group and Lie algabra
  structures at all levels of construction and representation, in
  order to qualify our use of the Hamiltonian as the generator of
  dynamical time translations, and to preserve the underlying geodesic
  nature of the group of symplectic (=dynamical) transformations.}
\item{We made use of the geodesic structure in the identifications
  generator=tangent vector to a geodesic=derivation and the geometric
  notions of covariant derivative.  Thereby, dynamical evolution on
  phase space is represented by geodesic transport in the space of
  probability amplitudes, etc.}
\end{itemize}

In the third installment~\cite{III}, we demonstrated the existence of
things we expect to find in hamiltonian dynamics were properly
included  in our probabilistic representation of hamiltonian dynamics:
\begin{itemize}
\item{We demonstrated the connection between chaos and analytic
  continuation, e.g., how Devaney chaos should be an expected result emerging
  from our correlated dynamics, and in its representation by
  probability amplitudes which are both analytically continued.}
\item{Because the symplectic group is geodesic and has hyperbolic
  generators, we expect the existence of fractals in both our dynamics
  on phase space and in our dynamically evolving probability
  amplitudes.  Fractals are typically generated by hyperbolic affine
  transformations. Our resonances are associated with a Julia set,
  and are generated by the hyperbolic generators of a geodesic semigroup.}
\item{In non-linear dynamics language, the equilibrium state which a
 resonance evolves towards would be called a strange attractor.}
\item{We demonstrated explicitly the role of resonances in a weak, or
  local, version of the Second Law of Thermodynamics, generic to all
  non-compact dynamics.  We argued it may be possible to arrive at a
  global, or strong, for of this law as well, also predicated on the
  presence of resonances.}
\end{itemize}

In this fourth installment, we have:
\begin{itemize}
\item{Looked for a well defined covering structure for what has gone
  before.}
\item{For the islands of stability, we have found this in the unitary
   Clifford algebras.  }
\item{For the generators of resonances, there are orbits of
  transformations (and their representations) for which we have
  offered no mathematically secure construction, although there is
  grounds for optimism.  There should be a sort of symplectic Clifford
  ``semialgebra'' for the representation of all of dynamics,
  reflecting  a semigroup of symplectic transformations, etc., rather
  than a full group structure, etc.}
\item{Demonstrated that there appears to be something akin to a rigged
  Hilbert space structure in the nuclear part of the unitary Clifford
  algebras for phase space,
making the rigged Hilbert space structure of our representation spaces
  appropriate.
(A similar structure exists for the nuclear symplectic Clifford algebras.)
We also demonstrated an enormous diversity of seemingly well behaved
mathematical structure associated with our unitary Clifford algebras.}
\item{We showed, as a mathematical truth, that irreversibility is
  intrinsic to dynamics, and existence of irreversibility depends only
  on the presence of correlation in the {\em continuous dynamics} of a
  dynamical system.  Without correlation, there is no continuous
  (geodesic) hamiltonian dynamics on phase space, unless it be
  arbitrarily postulated.  We have seen that, in general, reversible
  dynamics does not exist if one relies upon the geometry of phase
  space only. One must postulate away {\em topological obstructions 
  otherwise possible} if one wants continuous invertible dynamics
  in the general case. This is a mathematical confirmation of
  long held physical intuitions of the late Professor Prigogine when viewed in 
  conjunction with the results of installment three~\cite{III}, where we showed
  the presence of fractals and that resonances were evolving {\em toward}
  towards a strange {\em attractor} - the new equilibrium that a resonance
  evolves towards is an attractor so that there is an arrow to dynamical
  evolution.}
\item{Showed how bosonic and fermionic representations of dynamics
  emerge from alternative topological completions of the same set.  We
  argued that bosons are associated with the finer topology, and so
  seem more intimately associated with waves, while fermions are
  associated more intimately with the particle nature of the
  structures under study.  The wave--particle duality was explored.}
\item{Used the geodesic structure of the symplectic transformations to
  demonstrate the existence of Lie algebra valued connections, and
  used this in turn to define a gauge theory as to the islands of
  dynamical stability, showing a dynamical origin to the unitary gauge
  group.  Is was argued that this mathematically well defined unitary
  core may be extended to the orbit of the semigroups of symplectic
  transformations spanning the remainder of the structure of possible
  dynamics.  (Mathematical optimism at work.)}
 \item{In our construction, it would seem that the gauge freedom prevously
  dealt with by contraints, multipliers, etc., has been tacitly dealt
  with by stability considerations.  (Mathematical optimism at work.)}
\item{We adopted a field theoretic interpretation for the
  constructions and explored the generic span of compact hamiltonian
  dynamics for various numbers of fields.  As a gauge field theory,
  we were dealing with
  the potentials of the fields rather than the fields directly.  Lack
  of manifest covariance is dealt with by covariant associations due
  to isomorphism.  Particles are associated with the islands of stability
  of the potentials of the fields.}
\item{Particles are minimum uncertainty eigenstates, and the
  transformations between particles are analogous to the dynamical
  squeezings of the electromagnetic field in quantum optics.}
\item{In the case of three fields, we recover the Standard Model
  spectrum exactly, although with some differences in
  interpretation.  There are modifications to the mathematical
  understanding of bosons and fermions, and terms such as ``quark
  gluon plasma'' are mathematically found to be lacking in precision,
  although there may be some underlying associations between the
  fermionic and bosonic representations which make this and similar
  phrases physically meaningful in a qualified sense.}
\item{In the case of four fields, we found a strong connection to
  general covariance and hamiltonian General Relativity, but only so
  far as the islands of stability were concerned.}
\item{From the four field associations of our stable hamiltonian dynamics
  with the Lorentz group, we have deduced the existence of covariant
  associations for the characterizing quantum numbers for all
  elementary particles (e.g., whose quantum numbers are determined
  from a unitary gauge group) in the case of four fields.  All
  observers should agree as to the indentity of an elementary particle
  if there are four or fewer fields in nature.}
\item{We have shown that the mere existence in nature of non-compact
  dynamics demands the inclusion (somehow) of particles into our
  hamiltonian field dynamics which are not ``elementary''--not derived
  from a unitary gauge group--and which are chimeric in that they may
  be characterized differently by different observers: whose
  characterization is, in short, non-covariant.  This is demanded by
  the mathematics if we suppose that non-compact dynamics exists
  anywhere in nature, and if the universe is finite.}
\end{itemize}

Hopefully, this will help focus in the mind the largely generic
dynamical--mathematical approach which has led us to this seeming
dichotomy between dynamics and general covariance.  There are specific
predictions which may be anticipated as stemming from the spectrum
generating algebra approach to our four field theory, and so we have
an initial touchstone to test our dynamical description.  The
predictions as to the non-equilibrium aspects of dynamics are largely
inferential, given the lack of general covariance of the relevant
quantum numbers, and so specific inferences must be looked for if they
are to be accepted.

\appendix
\section{Basis Normalization and Squeezing}\label{sec:squeezing}

In Section~\ref{sec:spinors}, we saw how the creation and destruction
operators may be regarded as the canonical quantization of the real
Witt basis, just as the familiar position and momentum operators of
quantum theory may be thought of as the canonical (Dirac) quantization
of the classical position and momentum.  There are other operators
related to the creation and destruction operators and the position
and momentum operators (differing primarily in normalization) which
are especially useful for some of the more elementary considerations
of the squeezed and coherent states of quantum optics.

Taking the $A $ and $A^\dag$ as in installment two~\cite{II}, define
\begin{eqnarray}
\hat{a_1} &= \frac{A+A^\dag}{2} \nonumber \\
\hat{a_2} &= \frac{A-A^\dag}{2i}
\end{eqnarray}
and then define the new normalization of the position and momentum
operators as:
\begin{eqnarray}
\hat{x} &= \sqrt{2} \hat{a_1} \nonumber \\
\hat{p} &= \sqrt{2} \hat{a_2}  \quad .
\end{eqnarray}
Then $[ \hat{x} ,\hat{p} ] =i\II $, eleminating the $\hslash$ from the
familiar commutation relation of position and momentum--we could
obviously change the commutator to the identity by further alteration
of the normalization.  One of the simplest squeezing operators
is~\cite{dutra}:
\begin{equation}
Z= - \frac{1}{2} (\hat{x}\hat{p} + \hat{p}\hat{x} ) \quad  ,
\label{eq:squeezeop}
\end{equation}
The full gamut of squeezing operators belong to the relevant symplectic
Lie algebra, here the ``single photon'' lie algebra $\mathfrak{sp} (2,\R )
\cong \mathfrak{su} (1,1) \cong \mathfrak{o}(2,1)$.  Just as there
are one and two photon squeezed states in quantum optics, there should
be analogous one and two field squeezing operators, etc., in hamiltonian
field dynamics.

The traditional treatment of squeezed states has been in terms of
real and imaginary parts of a wave function and not of $p-q$ quadratures.
Note there is a symplectic
structure in either situation, and the present paper follows the more
ecent trend of thinking in terms of $p-q$ quadratures and a rotating
$(p,q)$ space ellipse being squeezed, etc.

The gist of this is that there should be multifield analogues to all
the squeezing phenomena of quantum optics if the present mathematical
development is to be believed.


\begin{thebibliography}{00}
\bibitem{whitney}
{If space $E$ has gauge froup $G$, then $E\oplus E$ has gauge group
isomorphic to $G\times G$.}
\bibitem{note0x5} 
{The full symplectic semigroup--the semigroup of dynamical
  transformations--is of course larger than the unitary group which is
  our gauge group.  Thus, the full scope of our dynamics is larger
  than the span of the gauge group structure,  It is not unreasonable,
  then, to speak of dynamical symmetry reduction.  But, the term
  ``spontaneous symmetry breaking'' is essentially meaningless, and
  deprecated in our formalism.  What actually occurs is merely
  reduction of the principal fiber bundle whose group is a proper
  subgroup of the  full dynamical semigroup.  It is a representation
  of  a dynamical process.}
\bibitem{note0x7} 
{If so, our hamiltonian formalism is largely generic, while the
  Lagrangian approach to field theory has been fine tuned over
  decades.  If the two disagree, it seems probable that this would be
  the result of some non-generic element figuring significantly in the
  Lagrangian approach.  Nature may or may not allow faithful
  description by generic means.}
\bibitem{note0x99} 
  {Rather that using for the symplectic form the matrix $J= \begin{pmatrix}
  0 & \II \\ -\II & 0   \end{pmatrix}$, one might use $J'= \begin{pmatrix}
  0 & g \\ -g & 0   \end{pmatrix}$, for instance.  This would allow
  exploration of the
  actual quantization of spacetime itself by respecting the spacetime
  metric as applied
  to the canonical position and momentum variables.}
\bibitem{note0x100}
  {There are associations between an affine frame bundle over a
  relativistic  world manifold and $\R^4$, since its structure group
  is  reducible to the linear group $GL_4$.  See, e.g., section 4.1
  of~\cite{sardanashvily}.   On this basis, there is grounds for
  speculation that our  ``generic'' results have in fact some sort of
  covariant  associations, but to extract anything well defined from
  this tenuous link is not trivial.  If this speculation follows, it
  would seem that our generic results should apply regardless of whether
  or not there is any Planck length acting as a fundamental length to
  spacetime.  By this, we therefore suggest speculatively that our
  generic approach is ``strongly generic'' and independent even of
  the precise nature of spacetime.}
\bibitem{note1} 
{According to O'Raifeartaigh~\cite{oraifeartaigh}, the standard
Model is more properly
$S\left( U(3)\times U(2)\right)$, with unbroken $U(3)=
SU(3)\times U(1)/Z_3$.  $SU(4)$ figured in the first GUT
proposal~\cite{pati1,pati2}.
See also~\cite{rajpoot}.  $SU(4)\cong Spin (6)$ can be thought of as
the compact version of
$SU(2,2)\cong Spin(4)$, the twistor group or the conformal group which
describes gravity.

There may well be some connection of a more concrete nature
with the twistor formalism of Penrose, given our
(momentum spinor)$\times$(position spinor) structure in the original
representation of phase space.}
\bibitem{note2} 
{It is interesting to speculate whether the use of a seminorm
topology and resultant semigroups is also necessary mathematically, and
this may be taken up by the author at a later date.  The importance of
initial conditions has been emphasized by R. Peierls and his
school~\cite{peierls1,peierls2}.  It is known mathematically that the
complete specification of initial and boundary conditions results in
semigroups of evolution~\cite{evolution}.
At the very least, our use of seminorm and resultant semigroups here
provides us with a vehicle to receive the complete specification of
dynamical initial and boundary conditions.  Here, the specification would
be probabilistic and not for the point localizations on phase space of
classical mechanics.}
\bibitem{note3} 
{This isomorphism is unique up to an
equivalence.  If there are isomorphisms from some aspect of our
dynamical structure into two separate $Cl_{O(1,3)}$ algebras, then
their substructures are isomorphic to each other not only because of
the isomorphism of algebras, but also
by way of their isomorphism to the geometric
algebra of the canonical coordinate space, and so any alternative
structures can differ by only the choice of frame used to describe
them.}
\bibitem{note4} 
{We can treat the space as if it were obtained by lifting direct sums
  of independent states, e.g., number states of a special sort, for
  which both creation and destruction operators are continuous.  The
  existence of such states is known.  See chapter 2
  of~\cite{dirackets}.  }
\bibitem{note5} 
{One of the key ingredients in making extended dynamical structures
  well defined, e.g., beyond the island of stability associated with
  the unitary transformations and their orbit, is the fact that the
  symplectic (semi-)group is geodesic as to each locally compact
  sheet, meaning that its Lie algebra may be associated with
  connections.  Thus, in electromagnetism the relation $\mathbf{B} =
  \nabla\times\mathbf{A}$ is undefined at the singularity of a point
  charge if functional analysis considerations are all that prevail,
  but when viewed geometrically, with $\mathbf{B}$ as the curvature of
  the connection $\mathbf{A}$, this relation is well defined.  This is
  well understood in the context of non-trivial bundles, and is
  the reason that gauge theoretic approaches have found much currency
  in modern mathematics, but it is the geodesic geometry and not the
  group structure (and well defined classical homotopy) which seems to
  be the governing factor here: a full blown gauge theory seems to
  assume more structure than is needed for a well defined affine
  (i.e., geodesic) geometry.  There are deep mathematical issues here
  which we will merely express an optimistic view of at this time,
  since they are peripheral to our necessary concerns.}
\bibitem{note6} 
{Since we have
a non-trivial spacetime around,
presumably there is a representation of the inhomogeneous
Lorentz group available somewhere which meets Wigner's definition of a
particle.}
\bibitem{note8} 
{As an interesting aside, note the complexification of the real
representation on $\F$ effectively results in a doubling of Hilbert
spaces.
If the real Lie algebra $\g$ generating $G_0$ is represented on $\F_0$
of the Gel'fand triplet $\F_0 \subset \HH_0 \subset \F^\times_0$,
then the analytic continuation $\g\longrightarrow \g^\C$ generating
$G^\C$ represented on $\F$ of the Gel'fand triplet $\rhs$ can
be seen by the preceding
argument as leading to the representation of a group isomorphic
to $G_0 \times G_0$ on $\HH$, e.g., looking like ``$G_0 \oplus G_0$''
represented on ``$\HH = \HH_0 \oplus \HH_0$''.  This looks reminiscent
of the Liousville space formalism, and there is perhaps some
connection.}
\end{thebibliography}
\end{document}